\documentclass[a4paper,10pt,3p,final,pdftex]{elsarticle}
\pdfoutput=1

\biboptions{sort&compress}

\usepackage{graphicx}
\usepackage{amssymb}
\usepackage{amsmath}
\usepackage{amsfonts}
\usepackage{array,hhline,dcolumn} 
\usepackage{epstopdf}

\usepackage{algorithm,algorithmic}
\usepackage{url}
\usepackage{color}
\usepackage{tikz}
\usepackage{multirow}

\newdefinition{remark}{Remark}

\newcommand{\opal}{\textsc{OPAL}}
\newcommand {\htp}{$\text{H}_2^+$}
\newcommand {\hthreep}{$\text{H}_3^+$}
\newcommand {\hmi}{$\text{H}^-$}

\newcommand {\htpe}{$\text{H}_2^{+*}$}

\newcommand {\hze}{$\text{H}_0$}

\journal{Version \today }

\begin{document}

\begin{frontmatter}

\title{Multimegawatt DAE$\delta$ALUS Cyclotrons for Neutrino Physics}

\author[iba]{M. Abs}
\ead{michel.abs@iba-group.com}

\author[psi]{A. Adelmann\corref{cor}}
\ead{andreas.adelmann@psi.ch}

\author[mit]{J.R. Alonso}
\ead{jralonso@lbl.gov}

\author[mit]{W.A. Barletta}
\ead{barletta@mit.edu}

\author[huu]{R. Barlow}
\ead{R.Barlow@hud.ac.uk}

\author[infn-lns]{L. Calabretta}
\ead{calabretta@lns.infn.it}

\author[mit]{A. Calanna}
\ead{alessandra.calanna@gmail.com}

\author[mit]{D. Campo}
\ead{dcampo@lns.infn.it}

\author[infn-lns]{L. Celona}
\ead{celona@lns.infn.it}

\author[mit]{J.M. Conrad}
\ead{conrad@MIT.EDU}

\author[infn-lns]{S. Gammino}
\ead{gammino@lns.infn.it}

\author[iba]{W. Kleeven}
\ead{Willem.Kleeven@iba-group.com}

\author[umary]{T. Koeth} 
\ead{koeth@umd.edu}

\author[infn-lnl]{M. Maggiore}
\ead{mario.maggiore@lnl.infn.it}

\author[rik]{H. Okuno}
\ead{okuno@riken.jp}

\author[infn-lnl]{L.A.C. Piazza}
\ead{piazza@lns.infn.it}

\author[psi]{M. Seidel}
\ead{mike.seidel@psi.ch}

\author[col]{M. H. Shaevitz}
\ead{shaevitz@nevis.columbia.edu}

\author[psi]{L. Stingelin}
\ead{Lukas.Stingelin@psi.ch}

\author[mit]{J. J. Yang}
\ead{Jianjun.Yang@psi.ch}

\author[ice]{J. Yeck}
\ead{jim.yeck@icecube.wisc.edu}

\address[umary] {Institute for Research in Electronics and Applied Physics, University of Maryland, College Park, Maryland, 20742}
\address[psi]{Paul Scherrer Institut, CH-5232 Villigen, Switzerland}
\address[mit]{Department of Physics, Massachusetts Institute of Technology}
\address[col]{Columbia University}
\address[infn-lnl]{Istituto Nazionale di Fisica Nucleare - LNL}
\address[infn-lns]{Istituto Nazionale di Fisica Nucleare - LNS}
\address[rik]{RIKEN, Nishina Center for Accelerator-Based Science, Hirosawa, Wako 351-0198, Japan}
\address[huu]{Huddersfield University, Queensgate Campus, Huddersfield, HD1 3DH, UK}
\address[ice]{IceCube Research Center, University of Wisconsin, Madison, Wisconsin 53706 }
\address[iba]{IBA-Research }
\cortext[cor]{Corresponding author}

\begin{abstract}
 DAE$\delta$ALUS (Decay-At-rest Experiment for $\delta_{CP}$ studies At the Laboratory for Underground Science) provides a new approach to the search for CP violation in the neutrino sector. High-power
 continuous-wave proton cyclotrons efficiently provide the necessary proton beams with an energy of up to $800$ MeV to create neutrinos from pion and muon decay-at-rest. 
The experiment searches for $\bar{\nu}_{\mu} \rightarrow \bar{\nu}_e$ at short baselines corresponding to the atmospheric $\Delta m^2$ region. The $\bar{\nu}_e$ will be detected via inverse beta decay. Thus, the cyclotrons will be employed at a future ultra-large gadolinium-doped water or scintillator detector.

In this paper we address the most challenging questions regarding a cyclotron-based high-power proton driver in the megawatt range with a kinetic energy of $800$ MeV.\  Aspects of important subsystems like the ion source and injection chain, the magnet design and radio frequency system will be addressed.

Precise beam dynamics simulations, including space charge and the \htp\ stripping process, are the base for the characterization and quantification of the beam halo--one of the most limiting processes in high-power particle accelerators.
  
\end{abstract}


\end{frontmatter}

\nocite{*}
 
\section{Introduction}
\label{sec:intro}

The DAE$\delta$ALUS Collaboration is designing advanced
cyclotrons that accelerate molecular hydrogen ions to produce
decay-at-rest neutrino beams in a novel search for CP violation in
the neutrino sector.  The required cyclotrons must exceed the
performance of the world's current best-performing cyclotron, at the Paul Scherrer Institut (PSI), in both
energy and power, while, at the same time, maintaining cost-effectiveness.

We begin by discussing the experimental context for the machines,  providing the requirements which constrain  the design  described in this paper. The next section addresses the options for ions accelerated, leading to the identification of \htp\ as the most promising candidate. Subsequent sections address individual subsystems of the accelerator complex, giving performance requirements and our  design concept for meeting these requirements.  Designs are generally based on existing accelerator facilities, so necessary  extrapolations are anchored on demonstrated performance.  Areas are identified where technical innovations might possess elevated risks, and programs are described to reduce or mitigate these risks.

This paper addresses only the accelerator systems necessary to produce the beams of the required power and energy; the high-power target/dump systems required for pion production are recognized as requiring substantial physics and engineering design, and will be addressed in subsequent studies. A cyclotron-based system is studied here, as we are convinced that the compactness and most probably substantially lower cost of cyclotrons (over linacs) are telling advantages.

\subsection{Goals of the project}

The physics community has placed the search for evidence for
CP violation in the neutrino sector at the highest priority
\cite{APSNeutrinoStudy, NUSAG, P5}.  
CP Violation in the light
neutrino sector would manifest as a difference in the oscillation
probability for neutrinos versus antineutrinos.  This is a key piece
of evidence for the theory of leptogenesis \cite{leptogen1,leptogen2}.  
The search
is also motivated by the experimental observation of CP violation in
quarks.  Why would the quark sector show CP violation if the lepton
sector does not?  What would this say about quark-lepton unification?

\begin{figure}[ht!]
\begin{center}
{\includegraphics[width=2.4in]{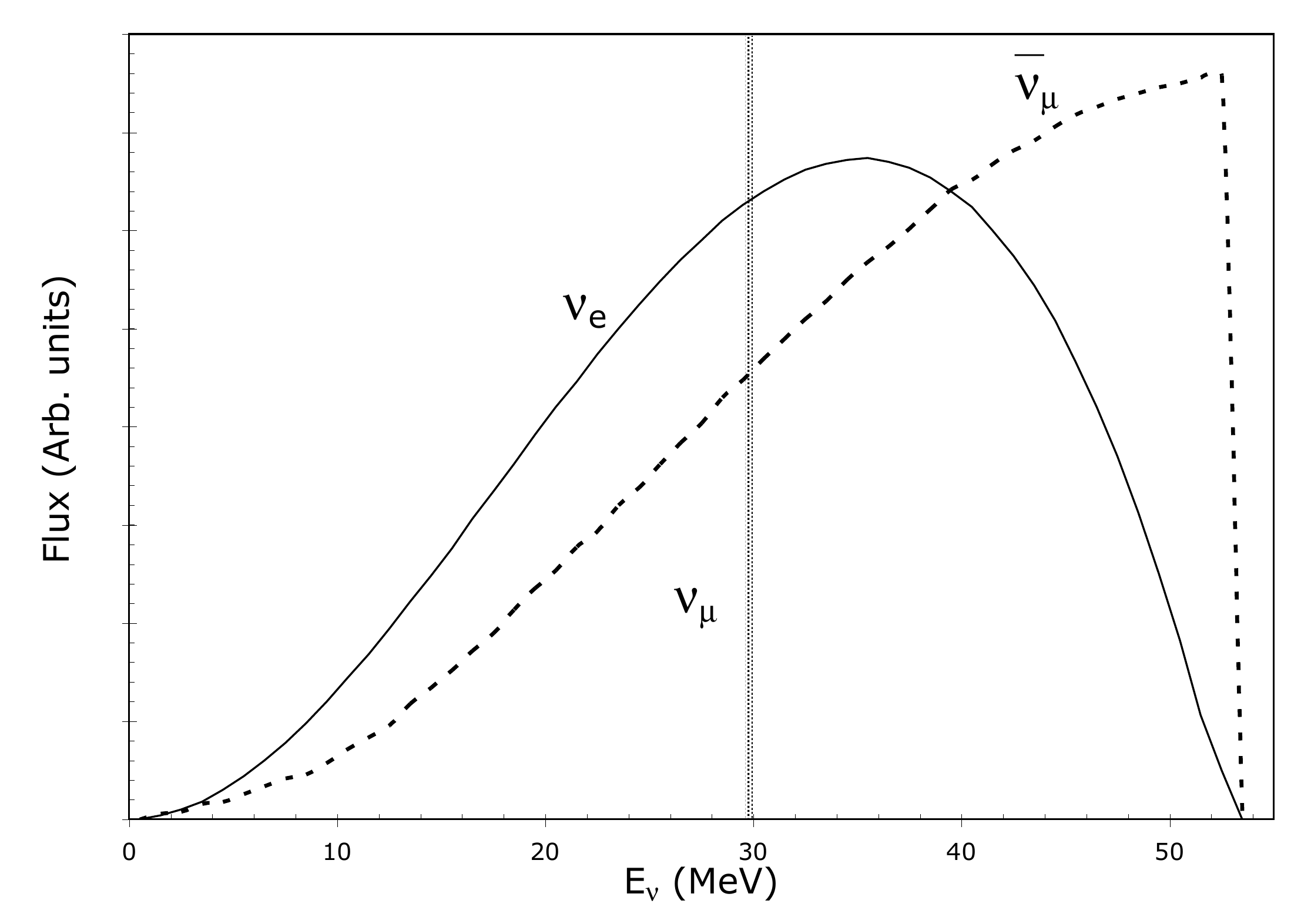}}
\end{center}
\caption{The energy distribution for a decay-at-rest flux.
\label{fig:cutoff}}
\end{figure}

The goal of the DAE$\delta$ALUS experiment is the search
for a nonzero CP violation parameter, $\delta$. 
The signal is  observed in $\bar \nu_\mu \rightarrow \bar
\nu_e$ oscillations.   The flux is provided by 800 MeV 
protons impinging on a carbon target
producing pions from the $\Delta$ resonance.  These come to rest in
the target and subsequently decay via the chain:
$\pi^{+}\rightarrow\nu_{\mu} \mu^{+}$ followed by $\mu^+ \rightarrow
e^{+}\bar{\nu}_{\mu}\nu_{e}$.  The resulting $\bar \nu_\mu$ flux is peaked and cuts
off at $52.8$ MeV as shown in Fig.\ \ref{fig:cutoff}.
 It is isotropic with a well-known energy dependence
for each of the three flavors, driven by the weak interaction.  Because almost all $\pi^-$ capture
before decay, the $\bar \nu_e$ fraction in the beam is expected to be $\sim 4\times
10^{-4}$ of the other neutrino flavors.   

This CP violation signal is extracted by measuring the oscillation 
wave as a function of distance ($L$) the neutrinos have travelled, as  illustrated in Fig.\ \ref{layout}.   The
DAE$\delta$ALUS design calls for measurements at three values:
$L=$ 1.5 km, 8 km, and 20 km.\ Three accelerator complexes, called ``near, mid, and far site," are used,
with the neutrinos impinging on a single ultra-large detector, which
may contain gadolinium-doped water or scintillator. In order to connect
the events to the accelerator site where the neutrinos were produced,
the sources must be run in alternating time intervals.   Also, roughly
50\% of the run must be accelerator-off to allow measurement of 
beam-off backgrounds in the detector.      DAE$\delta$ALUS requires 
average powers of 0.8, 1.6, and 4.8 MW from each site.   The duty factors and
the instantaneous power may be varied to achieve this.    

The average power of each station was chosen such that
DAE$\delta$ALUS matches the sensitivity of the 2011 design of 
the Long Baseline Neutrino Experiment (LBNE), a conventional neutrino beam
experiment, to CP violation \cite{LBNE}.    However, neutrino beams of this type suffer from high
systematic errors in this search. The DAE$\delta$ALUS multi cyclotron
design offers a statistics-limited alternative approach.   When
DAE$\delta$ALUS data is combined with the conventional beam data in
neutrino mode, the result is better than the expectation of a
conventional beam with a intensity upgrade, like Project-X \cite{EOI}. Thus,
this project has a very high discovery potential and is an exciting
option for the particle physics community.

\subsection{Layout of the proposed facility and basic parameters}
\begin{figure}[ht!]
\begin{center}
{\includegraphics[width=4.in]{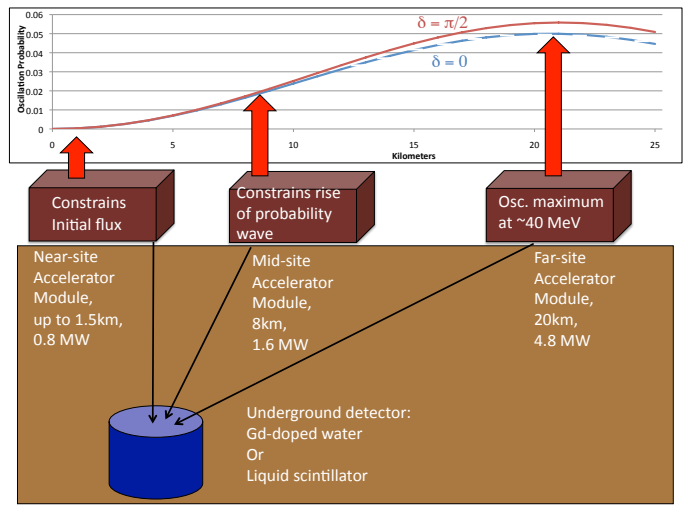}}
\end{center}
\caption{Schematic of the layout of DAE$\delta$ALUS accelerator modules. The powers at the respective modules, are average values based on a 20\% duty cycle.
\label{layout}}
\end{figure}

The DAE$\delta$ALUS three-accelerator-station design \cite{EOI} is illustrated in Fig.\ \ref{layout}. 
The near station measures the
initial flux, the mid station is at half of oscillation maximum, and the
far station is at oscillation maximum.  
Thus the oscillation wave can
be traced as a function of $L/E$ ($E$ denoting the energy of the beam), allowing sensitivity to $\delta$, as
is illustrated in Fig.\ \ref{layout}. 

The simplest
and least expensive option is to construct identical accelerator units --- ``modules'' --- which
 can be run for different time periods or grouped to
provide the required power at each station. Hence we focus here on the design of a single accelerator module.

A module consists of four basic components: the ion source (IS) and low-energy beam transport (LEBT); the Injector Cyclotron (DIC); and the DAE$\delta$ALUS Superconducting Ring Cyclotron (DSRC). Each module will also have its own dedicated target/dump, which
must all be identical to ensure commonality of the neutrino sources.

As will be shown in section \ref{sec:H2vsP}, we have selected \htp\ as the ion to be accelerated.  The DIC accelerates the beam to 60 MeV/amu, and extracts it with a ``classical'' electrostatic septum system. The \htp\ beam is transported to the DSRC, injected, and accelerated to the top energy of 800 MeV/amu.  A foil stripper converts the \htp\ into two protons that follow an extraction trajectory and are removed cleanly from the accelerator for transport to the target/dump. 

The 800 MeV/amu energy is selected to be comfortably above the pion production threshold, enabling good thick-target yields 
of $\pi^+$, but low enough to minimize contaminating $\pi^-$ production and decay in flight of the negative pions, an undesirable background.

\subsection{Possible connections with research beyond neutrino physics}

The two accelerators, DIC and DSRC, that form the basis of a module,
have applications beyond this experiment, such as in industry.   

\subsubsection{Medical isotope production} 
The DAE$\delta$ALUS Injector Cyclotron can accelerate more than 5 mA of particles with
charge-to-mass ratio of  1-to-2, including H$_2^+$ and $\alpha$-particles,
up to 60 MeV/amu.  This is in the energy range commonly used for
medical isotope facilities.       Isotopes that are produced at these
energies which are useful to the medical industry include
$^{52}$Fe, $^{122}$Xe, $^{28}$Mg, $^{128}$Ba, $^{97}$Ru, $^{117}$Sn,
and $^{82}$Sr  \cite{cost-cycl-2005}.

Cyclotrons on the market from
IBA Cyclotron Solutions \cite{IBA} and BEST Cyclotrons \cite{BEST}
produce $\lesssim$ 1 mA  of protons at 70 MeV.  
At the China Institute of Atomic Energy (CIAE), a compact 100 MeV \hmi\ cyclotron is being assembled, designed for currents up to 500 $\mu$A \cite{CIAE}.
Other designs have been studied for delivery of 2 mA of 70 MeV protons \cite{cost-cycl-2005}.  
Our DIC will produce 5 mA of H$_2^+$, which is equivalent of 10 mA of protons --- a significant increase in the beam current.

\subsubsection{Subcritical reactors}
The  DIC-DSRC combination is a promising source for accelerator-driven systems (ADS)
\cite{Rubbia,Fietier}. They provide an attractive approach to treatment of waste from reactors by burning and transmutation long-lived isotopes to shorter-lived ones, significantly easing the long-term storage requirements for this waste.  Recent studies from the  Nuclear Energy Agency (NEA) and the Organisation for Economic Co-operation and Development (OECD) \cite{NEA2002,McIntyre} detail these possibilities. Specifically, heavy transuranic elements with half-lives of many tens of thousands of years can be transmuted to lighter isotopes with half-lives of 1000 years or less. In fact, the DAE$\delta$ALUS \htp\ concept originated from a cyclotron for ADS development \cite{Lucianofirstpaper,Lucianosecondpaper}.

Another angle of application is that of a subcritical core of fissile materials. Such a core, having a K factor less than 1, will be inherently ``safe" from a criticality accident.  Such a core may consist of material such as thorium which cannot be configured with a K greater than 1. However, in principle, a proton beam of sufficient energy and power could provide a neutron flux adequate to drive K to 1 for such structures thus producing a ``fail-safe" reactor because as soon as the beam goes away, the K drops below 1.  The power available from such a reactor depends on the proton power provided and the K factor without beam, so for instance a $K=0.95$ would provide a multiplication factor of 20, and a $K=0.98$, a factor of 50.  A 10 MW beam into a $K=0.98$ core could produce a reactor yielding 500 MW.  This would be more than adequate to provide the power to drive the accelerator system, and deliver substantial power to the grid \cite{thorium-2002}. It is important to deliver the beam power relatively uniformly over the volume of the core, and a proton energy of $800 -1000$ MeV would adequately bathe a core of several cubic meters with secondary neutrons.

\subsection{Phasing of DAE$\delta$ALUS:  science of the sub-projects}

The DAE$\delta$ALUS project is being developed in phases that
allow establishment of the viability of each stage of the system.   Phase
1, development and testing of the ion source, is underway, as described in
this paper.  Phase 2 involves a full-scale test of the DIC and
associated LEBT.  This opens interesting scientific opportunities that are now
under study,  as discussed here.   Phase 3 is the construction of the first full accelerator
module,  which opens a ``near and mid site accelerator'' physics program, also
discussed below.   Phase 4 is the full three-site plan for the
CP violation search described above.  In this section we focus on
the physics case for  Phases 2 and 3.

\subsubsection{An Isotope Decay-at-Rest (isoDAR) beam produced using the DIC}

The injector provides a current of 5 mA \htp\ at 60 MeV. As discussed
above, this has applications in medical isotope production.  Here we
consider its application to discovery science, as a driver for an
isotope decay-at-rest (IsoDAR) electron antineutrino source. The design
uses the protons from the DIC to generate a high neutron flux which 
impinges on $^7$Li, with 99.99\% isotopic purity, to produce
$^8$Li, which subsequently decays.   
When paired with an existing scintillator-based detector (approximately 1 kton), this
$\langle E_\nu\rangle=6.4$ MeV source produces inverse
beta decay (IBD) interactions ($\bar \nu_e +p \rightarrow e^+ + n$)
as well as $\bar \nu_e$-electron scatters.

A wide range of physics studies are
possible with this intense, flavor-pure source.   Ref.~\cite{isodar}
describes the unprecedented sensitivity to electron antineutrino
disappearance at $\Delta m^2\sim$ 1 eV$^2$, reaching $>10\sigma$ and
including the capability to distinguish between  one and two sterile
neutrinos models. The source also provides a unique ultra-large
sample of $\bar \nu_e$-electron scatters for Beyond Standard Model
tests.\ Also, one can search for new particles produced in the
target that electromagnetically decay in the detector.

\subsubsection{Physics from a first ``near accelerator''}

The first stand-alone accelerator module can
produce a decay-at-rest beam from stopped pions and muons for a unique 
short-baseline neutrino program, exploiting the well-defined flux
of $\bar \nu_\mu$, $\nu_\mu$ and $\nu_e$.  This module is likely to be the
``Near Accelerator'' for the CP violation studies, and hence will be
placed near an ultra-large gadolinium-doped water detector or a scintillator detector.
At the same time, it is likely to be in a location where additional
smaller detectors can be installed underground nearby.  The 
physics program of the ``Near Accelerator" can exploit both
arrangements. Here we provide two examples.

Ref.~\cite{AgarwallaConradShaevitz} describes how this source can be paired with an
experiment like LENA \cite{LENA} to study the signal for $\bar \nu_\mu
\rightarrow \bar \nu_e$ and $\nu_e$ disappearance at high $\Delta
m^2$.    This study is complementary to
IsoDAR, which searches for $\bar \nu_e$ disappearance, allowing 
for interesting tests of models of multiple sterile neutrinos,
including those with CP and CPT (Charge, Parity, and Time Reversal) violations.  For the appearance
channel, pairing with a LENA-like detector, provides a stringent test of the
LSND and MiniBooNE signal regions at $>5 \sigma$ and  of $\nu_e$ $>3 \sigma$. 

A proposal for a small, dedicated short-baseline experiment 
for the purpose of discovery of coherent neutrino scattering
\cite{Anderson:2011bi} is also being considered for the ``Near Accelerator''.   This would be located in an underground 
chamber near to the module.   Coherent elastic neutrino- and WIMP-nucleus (Weakly Interacting Massive Particles)
interaction signatures are expected to be quite similar, and so a next-generation ton-scale dark matter detector could discover
neutrino-nucleus coherent scattering, a precisely predicted Standard
Model process. The flux from the decay-at-rest beam 
was shown to be ideal when paired with
planned dark matter detectors for this search.
An extension of this concept is to measure active-to-sterile
neutrino oscillations with neutral current coherent neutrino-nucleus
scattering \cite{Anderson:2012pn}.  The design uses
multiple decay-at-rest targets, at different positions, driven by a
single DAE$\delta$ALUS accelerator to produce a neutrino flux that
impinges on a single, stationary dark matter detector.   This induces
coherent neutrino scatters.  Deviation from the expected $L$
dependence for the events would indicate Beyond Standard Model physics.

\section{\htp\ for the DAE$\delta$ALUS project}
\label{sec:H2vsP}
Obvious choices of ion species, to obtain high-power proton beams, are bare protons, $H^-$, or \htp.\ In our case, $H^-$ can be excluded because of Lorentz stripping at the upper energies in the very strong magnetic field required for a compact machine. For example, the TRIUMF top energy is 500 MeV, and the highest field in the 18-meter diameter cyclotron is only 0.5 T.  At 800 MeV, the peak field would need to be no more than 0.3 T, making for an impossibly large machine.  

Protons and \htp\ are both suitable candidates; in Table\ \ref{tab:p-vs-htp} key differences are identified, and a degree of risk is given with respect to the desired performance parameters.  The prime advantages of \htp\ are reduced space-charge effects at low energies (due to the presence of two protons for every charge) and the ability to extract the beam at high energy with a stripping foil, reducing the need for the clean separation between turns that is mandatory for bare protons. 

The space charge of the particle beam produces a repulsive force inside the beam, which
generates detuning effects. A measure of the strength of this effect, the so-called generalized perveance  is defined by
\begin{equation} \label{eq:per}
K = \frac{qI}{2\pi\epsilon_0 m \gamma^3 \beta^3},
\end{equation}
where $q, I, m, \gamma$ and $\beta$ are respectively the charge, current, rest mass and the relativistic
parameters of the particle beam \cite{reiser2008theory}. The higher the value of $K$,
the stronger the space-charge detuning effects.

According to Eq. (\ref{eq:per}), the space-charge effects for the 5 mA of \htp\ beam in the DSRC are equivalent to a 2.5 mA proton beam with the same  $\gamma$. Consequently, they are similar to the space-charge effects present in the 2.4 mA proton beam being accelerated currently at PSI.
We also note that the 5 mA \htp\ injected into the DIC has a similar $K$ value as commercially available cyclotrons used for the production of radioisotopes. 
Another degree of freedom, in order to reduce space-charge effects, is of course the energy, in particular the injection energy and acceleration voltage. In this specific situation, we increase the typical injection energy of the DIC from 30 keV to 70 keV. 

However, since there are many other factors, such as the external focusing and total number of turns in the cyclotrons, which have important impacts on the overall space-charge effects, we cannot get a clear picture without precise beam dynamics calculations.  In this paper, the space-charge effects are studied quantitatively by self-consistent 3D models implemented in the code Object Oriented Parallel Accelerator Library (\opal)\ \cite{opal}. The beam dynamics model is described in detail in \cite{yang-1, PhysRevSTAB.13.064201}. For the DSRC, we implemented a simple stripper model into \opal\ in order to study the complex extraction trajectories of the stripped protons. 

We have identified the main challenges of accelerating \htp : the requirement for a substantially larger machine due to the higher magnetic rigidity of the ions, the need for         a higher vacuum due to the greater probability of collisional dissociation of the ions with residual gas atoms (at least an order of magnitude better vacuum than required for protons), and the presence of loosely bound vibrational states in the \htp\ ions emerging from the ion source. 

\begin{table}
{\renewcommand{\tabcolsep}{0.9cm}}
\caption{Comparison of Protons vs. \htp\ as candidates for DAE$\delta$ALUS cyclotron beams} \vspace{+3mm}
\centering
\label{tab:p-vs-htp}
\begin{tabular}{ l l l l}
\hline   
 Requirements  &     Protons &     \htp  \\
\hline
Ion source: &  &  \\
   Current ($>$40 mA)  &  Demonstrated  &  Expected to be achievable \\
   Emittance ($\epsilon_n$  $<$ 0.3 $\pi$ mm mrad) &  &  \\
\hline
Ion source: species purity  &  Analyzer separates species   &  Analyzer separates species \\
 &  other than protons &  other than \htp  \\
 \hline
 Ion Source: & & Need to suppress higher \\
  Control of internal degrees  &  Not applicable & vibrational states of \htp; \\
  of freedom  &  & requires demonstration \\
 \hline
Injector Cyclotron (DIC): & High space charge is a challenging &  \\
 capture of adequate beam & problem in compact cyclotrons:  &  Will present similar challenges,\\
  &  good capture efficiency demonstrated & but lower perveance implies\\
   & in separated-sector machines & a simpler problem\\
   & (PSI Injector) \\
\hline
DIC: Extraction & Expected to be achievable  & Expected to be achievable \\
\hline
Ring Cyclotron (DSRC): & & \\
Injection	& Expected to be achievable & Expected to be achievable \\
\hline
DSRC:  Beam rigidity & K = 800 & K = 3200 \\
 &  Could use normal magnets  &  Requires large, high-field \\
  & & Superconducting magnets \\
\hline
DSRC:  Beam dynamics   & Resonances may limit    &  Dangerous resonances   \\
 & achievable current  &  appear to be avoidable \\
 \hline
 DSRC:  Extraction  &  Requires clean turn separation & Stripping extraction with foil \\
  & at high energy, very difficult & does not need clean turn \\
   & to achieve & separation, but requires control of  \\
    & & halo in the extraction channel \\
\hline
DSRC: rf & Need very high-power rf & Relatively straightforward \\
& to get high energy-gain/turn & No need for very high power \\
& to obtain clean turn separation & \\
\hline
DSRC: Vacuum  &   Requirements easily achieved  &  Requirements substantially more stringent   \\
& ($<$ $10^{-7}$ torr) & ($<$ $10^{-8}$ torr) to minimize \\
& & beam loss through gas collisions \\
\hline
DSRC: Beam loss concerns & Turn separation at extraction & Lorentz dissociation of \\
& &     unquenched vibrational states \\
& & Collisions with residual gas \\
& & Neutrals emerging from foil \\
\hline         
\hline\end{tabular}
\end{table}

\subsection{\htp\ Dissociation} 
\label{sec:dissoc}
\begin{figure}[ht!]
\begin{center}
{\includegraphics[angle=0, width=0.7\linewidth]{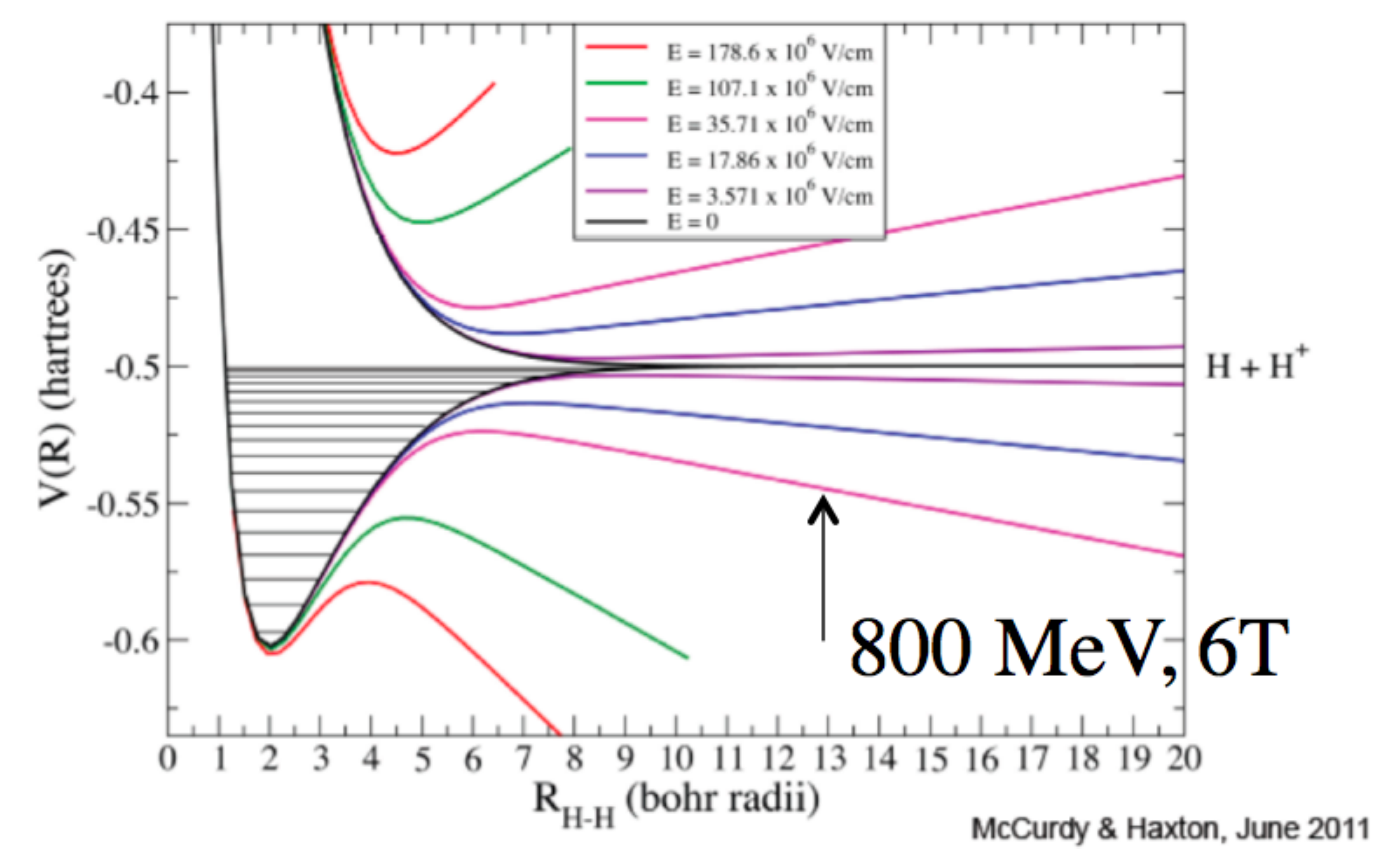}}
\end{center}
\caption{The \htp\ potential well, with vibrational states are indicated by horizontal lines. Distortion of the \htp\ potential well for an 800 MeV/amu ion in an electric field.  The distorted well results in shifted levels, however these are not shown.  The relativistic transformation of an 800 MeV/amu \htp\ in a 6 T magnetic field, resulting in an apparent electric field,  yields the distorted well shown in magenta, and noted with the arrow. \label{fig:disso2}} 
\end{figure}

The creation of \htp\ ions in the ion source results in the population of vibrational states.  As shown in Fig.\ \ref{fig:disso2}, the 17 bound vibrational states have binding energies from 2.7 eV (ground state) to essentially zero.  Because of molecular symmetry (implying lack of dipole coupling), the lifetime of these states is long compared to the residence time of the ion in the accelerator. The bound states result from different equilibrium distances for the protons bound by only one electron compared with two for the neutral molecule.  Figure \ref{fig:disso1} shows the Franck-Condon population distribution of these states, which is roughly independent of the source environment, e.g.,  plasma temperature \cite{disso-1}.  States whose binding energy is less than about 1 eV are likely to be Lorentz stripped in the 6T DSRC field at an extraction energy of 800 MeV/n; this fact is shown in
Fig.\ \ref{fig:disso2}.\ These ions contribute to beam loss.  As the total population of states of $n = 7$ or higher is roughly 10\%, the beam loss from this mechanism would be unacceptably high:  100 kW for 1 MW of extracted beam, which is a factor of 500 over the 200 watt ``loss budget" \cite{disso-3}. 

Initial experiments at the vibrational state detector of Oak Ridge National Laboratory (ORNL) are ongoing, and will characterize vibrational state populations from a representative ion source, and study in a separate ion trap the collisional dissociation process of \htp\ ions in helium gas.  This will be followed by later runs testing admixtures of noble gases in the ion source itself, and optimizing the source type and geometry to produce adequate currents of \htp\ devoid of the damaging vibrational states.

\begin{figure}[ht!]
\begin{center}
{\includegraphics[angle=0, width=0.8\linewidth]{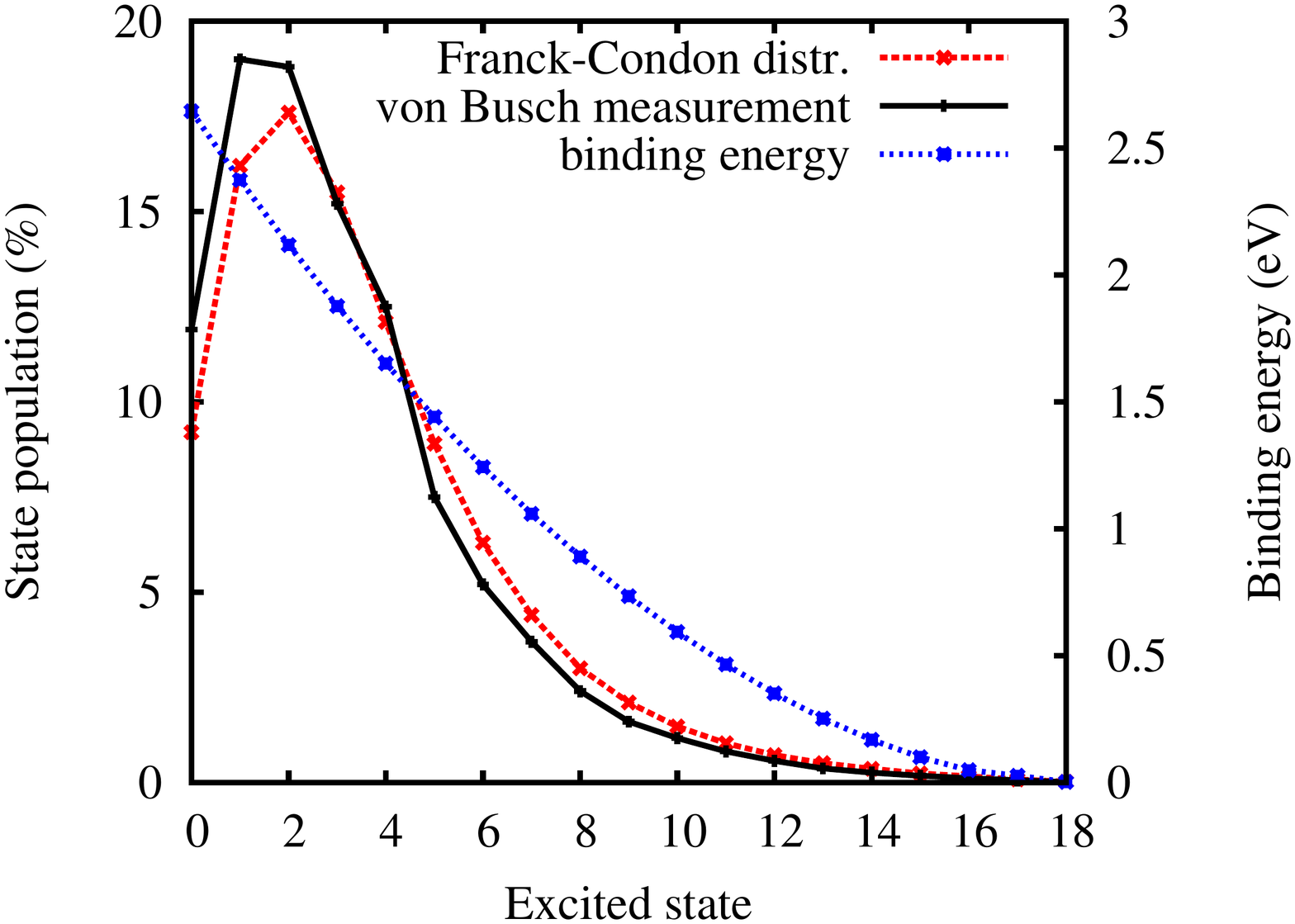}}
\end{center}
\caption{Vibrational state binding energies and populations for \htp\ ions \label{fig:disso1}}
\end{figure}

\subsubsection{Cross section of H$_2^+$ dissociation}
\begin{figure}[ht!]
\begin{center}
{\includegraphics[width=6.in]{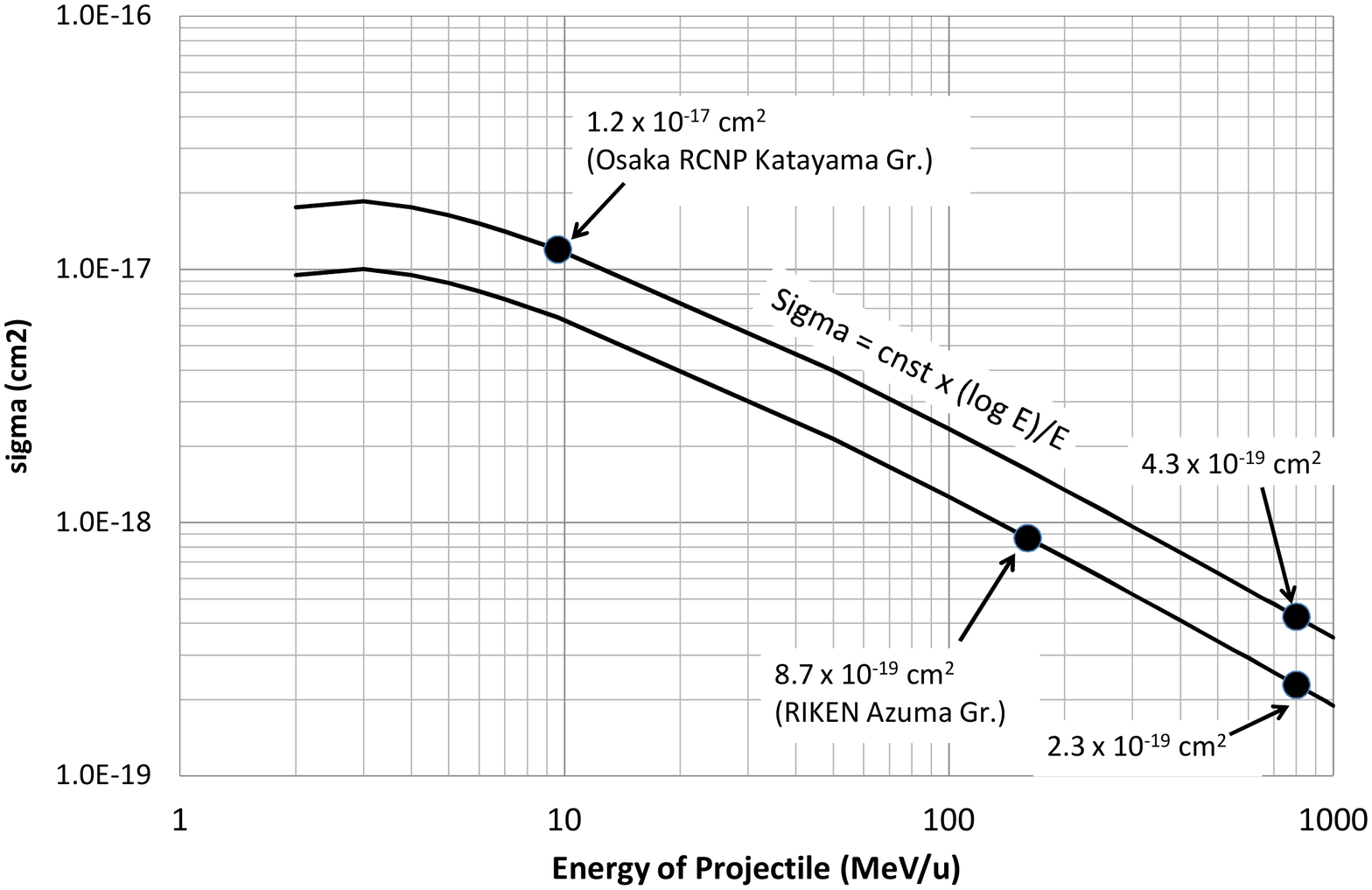}}
\end{center}
\caption{Energy dependence of the cross section of H$_2^+$  dissociation
\label{fig-crosssec1} }
\end{figure}

Survival rates of H$_2^+$ in C-foil were measured as functions of the foil thickness at 160~MeV/amu \cite{azuma} and at 9.6~MeV/amu \cite{katayama}. Cross sections of H$_2^+$ dissociation to two protons and an electron can be inferred from the measurements to be 8.7$\times$10$^{-19}$~cm$^2$ at 160~MeV/amu and 1.2 $\times$10$^{-17}$ cm$^2$ at 9.6~MeV/amu assuming that the channel to H$^+$ + H$^0$ is negligible. The cross sections can be extrapolated to be 4.3$\times$10$^{-19}$~cm$^2$ and 2.3 $\times$10$^{-19}$~cm$^2$ respectively at 800~MeV/amu using scaling law by log~(E)/E as shown in Fig.\ \ref{fig-crosssec1}.\ The two values at 800 MeV/amu suggest that survival rate at 800 MeV/amu on a carbon foil with a thickness of  2~mg/cm$^2$ is less than 1.0 $\times$ 10$^{-10}$. We conclude that 2~mg/cm$^2$ is thick enough to dissociate all the H$_2^+$ beam at the foil. 

The emission of H$^0$ after the foil is a possible point of concern because it will be an origin of uncontrolled beam loss. The break-up rate of the H$_2^+$ molecule into a proton and a neutral hydrogen is considered to be small. However, we have to take into account the recombination H$^+$ after the H$_2^+$ dissociation to neutral H$^0$ at the foil. Following \cite{schlachter}, the capture cross section of H$^+$ in the carbon foil is estimated to be 9.4 $\times$ 10$^{-34}$ cm$^{2}$. There is clear evidence that only a small amount of H$^0$ will survive; however, a dedicated experiment to measure the rate of H$^0$ is being planned.
 
\subsubsection{Vibrational state analysis} 
\htp\ ions emerging from a normal ion source will populate a range of vibrational states \cite{h2psrc}.  A technique for quenching loosely bound vibrational states has been suggested by Sen et al. \cite{disso-4}. This technique introduces noble gases $\text{H}e$ or $\text{N}e$ into the ion source because the collisional recombination reaction
\begin{equation}
		\text{\htpe}  +   \text{H}e    \rightarrow    \text{H}e\text{H}^+  +  \text{H}_0
\end{equation}
is exothermic for vibrational states with a quantum number higher than three.  It remains to be determined whether this mechanism can clean the beam emerging from the source of loosely bound states to better than the one part in 500 required to prevent the unacceptable beam losses, while still maintaining the high currents necessary to achieve the beam-power goals of the accelerator complex.

An experimental apparatus at ORNL, initially developed at CRYRING in Stockholm \cite{CRYRING93,Larsson93}, allows us to characterize vibrational state populations of low-energy molecular beams such as \htp\  \cite{disso-5,disso-6}.\ This detector bends the \htp\ ions through $90^\circ$ into the path of a 1 nm laser beam, with an overlap of about 18 cm length.  A few ions will be dissociated into a proton and a neutral hydrogen atom.  The beam, largely unaffected, is bent through another $90^\circ$ into a Faraday cup, but the few neutrals created will propagate through a thin foil, freeing electrons that are accelerated into a channel-plate detector for counting.  While difficult to calibrate, this detection scheme is suitable for establishing the presence or absence of states with binding energies less than about 1 eV.  Changing the wavelength of the laser light can probe down to different binding energies.  Assuming the raw beam extracted from a typical source follows the Franck-Condon distribution shown in \label{fig:disso1}, the ratio of neutrals counted to Faraday cup reading should correspond, for a 1 nm laser, to about 10\% of the beam.

The test stand also contains a stage where ions are slowed down and captured in a radio-frequency quadrupole (RFQ) trap, and can slowly drift through before being re-accelerated to enter the above-mentioned detector.  This trap can be filled with low-pressure gas, the primary purpose of which is to cool heavier-ion products.  The residency of ions in the trap is about a millisecond, substantially longer than the sub-microsecond transit time.  This system was developed for use with the Holifield Radioactive Beam Facility, but can be applied for the present studies by observing whether the vibrational states can be collisionally dissociated with an appropriate gas admixture.  This would be observed by a decrease in the neutral to Faraday-cup ratio in the detector.  Studying this ratio as a function of gas pressure and species is expected to provide insight into the collisional dissociation process in an environment independent from the very complex conditions present in an ion source plasma.

Results from the RFQ trap experiments will lead to testing of gas mixtures in the ion source plasma, and eventually to design of the appropriate source with the long confinement times required to achieve the 500-to-1 suppression of vibrational states.  A substantial R\&D program is envisioned to achieve this performance.
This experiment will lead to the specification and design for the DIC source capable of delivering the required very high current of \htp\ ions free of vibrational states.

\section{Ion source and Low-Energy Beam Transfer (LEBT)}
The ion source for DAE$\delta$ALUS is required to produce beams of \htp\ ions, free of weakly bound vibrational states, and an emittance suitable for axial injection into the DIC.\ For the near \& mid sites, the beam current out of the source must be sufficient to produce the $2$ mA average proton current at 800 MeV (assuming a 20\% duty factor).

To determine the required (\htp) current from the source, one must assess the overall efficiency of the transport and acceleration of the beam.  By far the greatest losses occur during inflection and capture in the DIC.  Once captured, the efficiency for transfer, acceleration, and extraction must be close to 100\%.

A practical value for the phase acceptance is $\pm 10^\circ$; hence, only approximately 
5\% efficiency can be expected. A longitudinal bunching system in the LEBT  improves the capture efficiency. Assuming a conservative bunching factor of two, an overall 
injection efficiency of 10\% is conceivable; hence, the source has to deliver an average \htp\ current of 50 mA. The source can be pulsed to accommodate the 20\% duty factor.  For optimizing the rf-system, the pulse length will be between 1 and 10 milliseconds, this leads to a reduction of the heat load on the source, allowing for higher instantaneous power, and therefore, higher beam currents.  

In principle, obtaining \htp\ ions is an easy extrapolation from conventional proton sources.  An analysis of beams from such a proton source will always contain fractions of \htp, \hthreep, and other positive molecular ions arising from contaminants in the source.  By adjusting operating parameters, the fraction of \htp\ can be improved; efficiencies as high as 80\% for \htp\ have been reported \cite{joshi-1}.  However, no characterization of the vibrational state populations has been performed, nor attempts made so far to suppress the weakly bound ones.

\subsection{Central region test}
An experiment is being designed to study the beam dynamics from the existing Versatile Ion Source (VIS), designed and build at INFN-LNS (Catania), up to the matching point of the spiral inflector. The aim of the test is to study inflection and capture efficiency of high-current \htp\ beams in the central region.

\begin{figure}[ht!]
\begin{center}
{\includegraphics[angle=-0, width=0.8\linewidth]{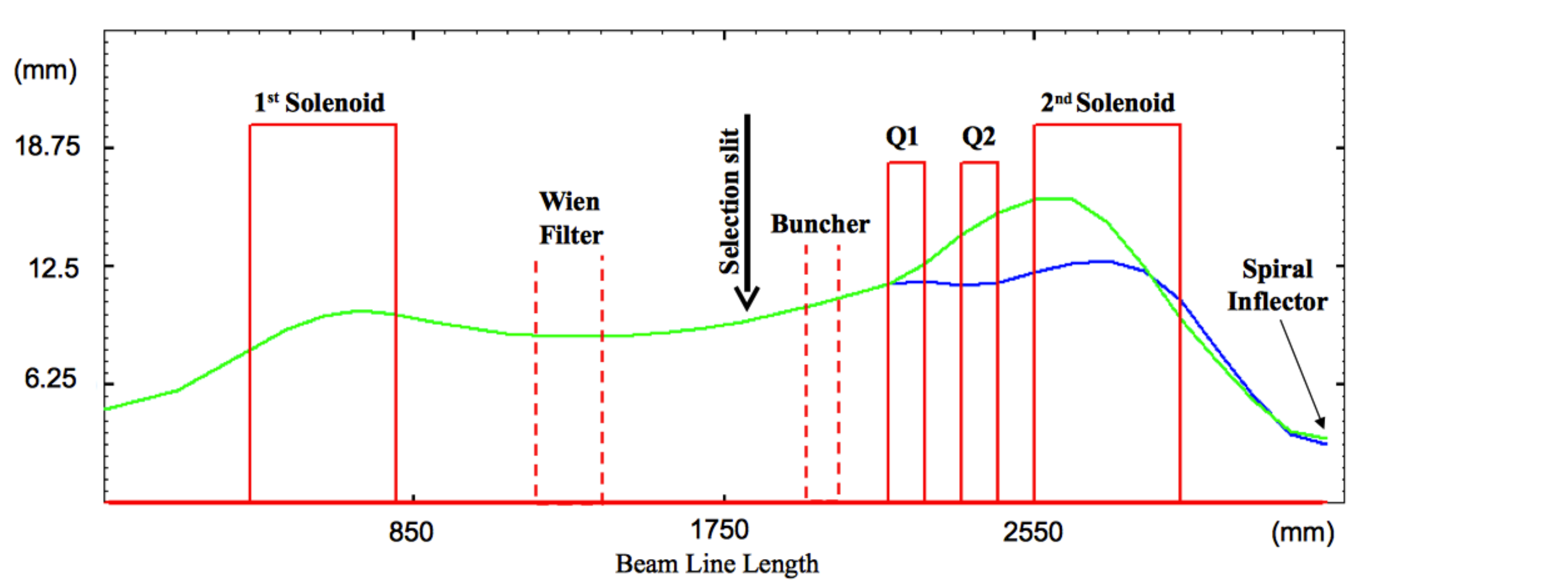}}
\end{center}
\caption{Beam envelope of the \htp\ beam along the injection line, from the ion source up to the spiral inflector of  the DIC
\label{fig:ch3-2} }
\end{figure}

Figure \ref{fig:ch3-2}  shows the beam envelope through the LEBT: a 40 cm long solenoid that focuses the beam through a collimator placed about one meter away, downstream of the Wien filter.  The Wien filter will be 15 cm long, with magnets producing a field of about 850 Gauss, and a crossed electric field of about 2 kV/cm
separating the components of the beam: consequently only Q/A = 1/2 ions have the correct velocity to pass through the filter undeflected.

At the location of the slits, 40 cm downstream, the proton and \htp\ peaks are separated by about 6 cm. Downstream of the selection box, two quadrupoles and a second solenoid provide the necessary matching to the central region of the test cyclotron. In the actual DIC, these elements will be located within the axial injection line. The use
of skew quadrupoles will add an additional degree of freedom for matching the beam into the inflector channel.

An rf-buncher, operating at the fundamental frequency, provides longitudinal compression and, hence, better capture efficiency. Space-charge forces will limit the effectiveness of this bunching, but the lower perveance of the \htp\ ion (compared to protons of the same velocity) is expected to be an advantage. 

While the operating parameters of this test (source extraction voltage, buncher and filter parameters, and inflector design) are not likely to be exactly those of the final design for the central region of the DIC, information gained from the tests is expected to provide baseline data that can be used to benchmark simulation codes that will be used to design the actual central region.  

\subsubsection{The versatile ion source}

The compact VIS source \cite{trasco-1,celona-1,maimone-1} is an off-resonance microwave discharge source operating at 2.45 GHz
using permanent magnets and a simplified extraction geometry. Extraction from the VIS source is accomplished with a four-electrode accel-decel configuration to control backstreaming of electrons and maintain high-brightness optics for currents up to about 40 mA of protons at maximal 80 kV. The source has already demonstrated good continuos wave (CW) proton currents ($>$35 mA) with normalized emittance below 0.2 $\pi$ mm-mrad. 

A preliminary test for extracting \htp\ ions from this source yielded 20 mA from an 8 mm extraction aperture.  Substantially higher currents can be obtained by increasing this aperture to 10 or 12 mm, while still maintaining an overall emittance within the expected acceptance of the cyclotron.  In addition, different methods for exciting the plasma are being explored that might increase plasma density and improve source performance.
 
\subsubsection{Bunching}

The ion source delivers a beam that is continuos, on the scale of the cyclotron rf-frequency.  However, the cyclotron will only accept particles arriving within a narrow phase window of $\pm 10^\circ$. This implies that only about 5\% of the beam from the source will be captured in the cyclotron. In the first phase of the project a buncher in fundamental mode, and later a buncher running at the third harmonic of the rf, is expected to improve the bunching capture efficiency by factors of two to three. For bunching related discussions in this paper we assume a conservative bunching factor of two.


\section{The DAE$\delta$ALUS injector cyclotron}
\begin{figure}[ht!]
\begin{center}
{\includegraphics[angle=-0, width=0.8\linewidth]{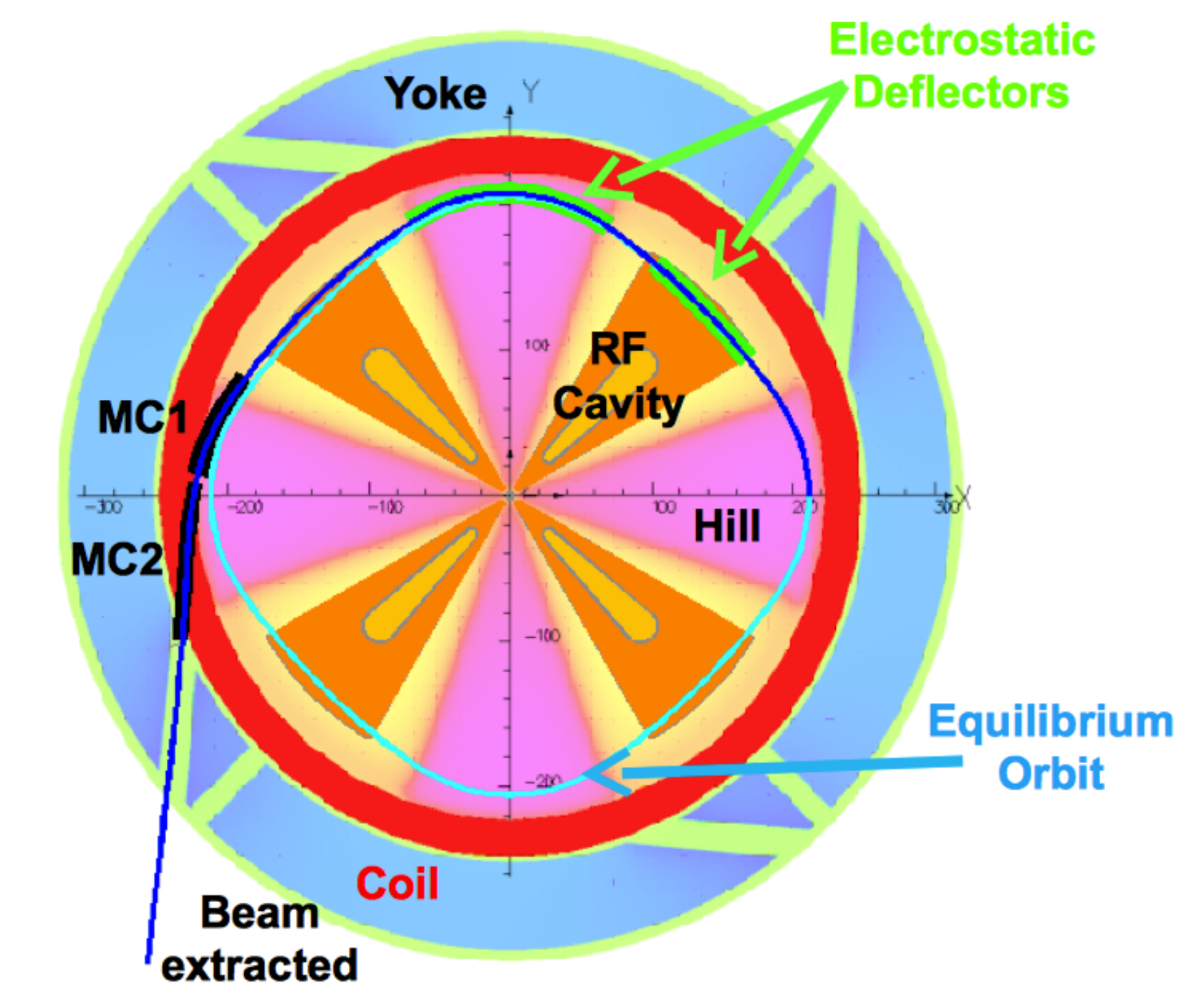}}
\end{center}
\caption{Layout of the DIC, a four-sector cyclotron with a pair of room temperature coils. The acceleration system consists of four rf-cavities that work at the frequency of 49.2 MHz at the 6th harmonic. The beam is injected into the center of the machine by using a spiral inflector, the extraction system includes two electrostatic deflectors and two magnetic channels. The extraction path of the \htp ions are shown, starting from the 0$^\circ$ azimuth.
\label{fig:ch3-4} }
\end{figure}

We study intensities between 1 and 5  mA of \htp\ up to the energy of 60 MeV/amu, but we are deferring the discussion of the spiral inflector and the central region to a
follow-up paper. A compact four-sector cyclotron is proposed and sketched in Fig.\ \ref{fig:ch3-4}. It is based on the parameters achieved by existing commercial compact machines \cite{kleeven-2011} and accelerators for research purposes.
Existing commercial \hmi cyclotrons use axial injection of the beam at 30 keV and can accelerate and extract beam currents of about 2 mA. In this design, the \htp\ beam is injected at about 35 keV/amu such that the generalized perveance is in the same range as for commercial cyclotrons. In this way, it is considered that the space-charge problems at injection will be equivalent to those of the proven commercial cyclotrons.
However, space-charge effects at low energy and the optimization of the extraction system remain the primary challenges; they are addressed in this section.

The magnetic field design has to satisfy requirements concerning beam dynamics, as well as the technical and economical feasibility of the machine.  These requirements can be listed, as follows:
\begin{itemize}
\item minimize the effects on the beam quality due to the resonance crossing and, in particular, to guarantee the 
conditions of vertical and radial stability ($\nu_r > 1$ and $\nu_z > 0.5$)
\item the average magnetic field  errors must be on the order of a few times $10^{-4}$ to ensure that the integrated rf-phase slip remains smaller than $20^\circ$
\item minimize the coil current density in order to reduce the power consumption and the heating of the pole
\item minimize the iron weight to reduce the cost, while an acceptable value of the stray field has to be less 
	     than 500 gauss at one meter away from the cyclotron iron yoke.
\end{itemize}

A large hill gap of 10 cm was chosen to have 
enough space for the beam transport and to achieve a good conductance for the vacuum system. A total of eight cryopanels are located inside the dee electrode 
of the rf-cavities. The vacuum system will be able to achieve better than $10^{-7}$ mbar, which is a necessary condition to reduce the beam losses 
due to \htp\ stripping. 

\begin{figure}[ht!]
\begin{center}
{\includegraphics[angle=-0, width=0.6\linewidth,trim=0.0cm 5.0cm 0.0cm 2cm]{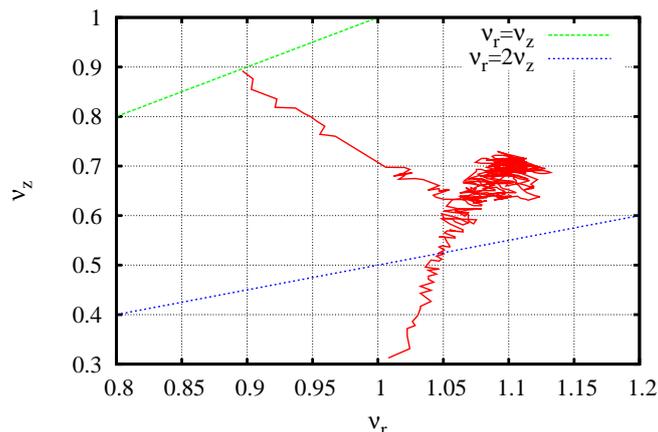}}
\end{center}
\caption{The betatron tune diagram in the DIC cyclotron
\label{fig:ch4-1}}
\end{figure}

\begin{figure}[ht!]
\begin{center}
{\includegraphics[angle=-0, width=0.6\linewidth,trim=0.0cm 7.0cm 0.0cm 7cm]{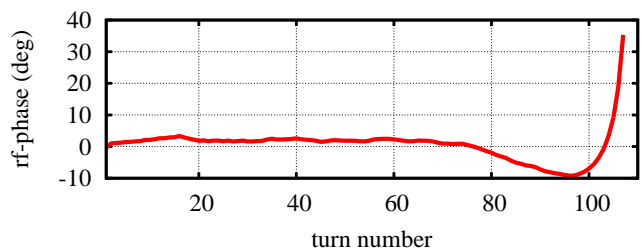}}
\end{center}
\caption{The integrated rf-phase shift in the DIC cyclotron
\label{fig:ch4-2}}
\end{figure}

The angular width of the hill, in the range of $28^\circ - 40^\circ$, optimizes magnetic field properties, resulting in a valley larger than $45^\circ$, allowing 
easily the installation of rf-cavities working at the 6th harmonic. According to our
preliminary model simulation, in the acceleration region, from 0.5 to 60 MeV/amu,
the isochronism accuracy is better than $5\times10^{-4}$. The betatron tunes are shown in Fig.\ \ref{fig:ch4-1}.
The \htp\ reaches  61.7 MeV/amu after 107 turns, maintaining an integrated phase shift in a narrow range of about $\pm 10^\circ$ (excluding the extraction
region), as is shown in Fig.\ \ref{fig:ch4-2}. 

Results from OPAL simulations show that space charge helps to from an approximately circular and stationary beam distribution in the horizontal-longitudinal plane. 
However, a beam halo on the order of $10^{-4}$ of the intensity is formed. In order to reduce the halo a  
four-collimator scheme was developed at turn six, which is around 1.9 MeV/amu. With the four-collimator scheme in place, simulations where carried out for 1 and 5 mA 
and two conceivable bunch lengths of $\pm 5^\circ$  and $\pm 10^\circ$ of the \htp\ beam. 
The results with respect to the beam losses are given in Table\ \ref{tab:ch4-6} at the extraction septum.
We conclude that, according to the experience of the PSI Injector II cyclotron, these values are acceptable.  

\begin{table}[hbt]
\centering
\begin{center}
\caption {The beam loss on the septum}
\label {tab:ch4-6}
  \begin{tabular}{cccc}   
	\hline
	Initial phase width  & 1 mA & 5 mA \\ 
	\hline
          $\pm 5^\circ$  & 6 W& 24 W \\
	$\pm 10^\circ$  & 14 W& 22 W \\
	\hline
  \end{tabular}
\end{center}
\end{table}

\begin{figure}[ht!]
\begin{center}
{\includegraphics[angle=-0, width=0.8\linewidth,trim=2.0cm 6.0cm 0.5cm 6cm]{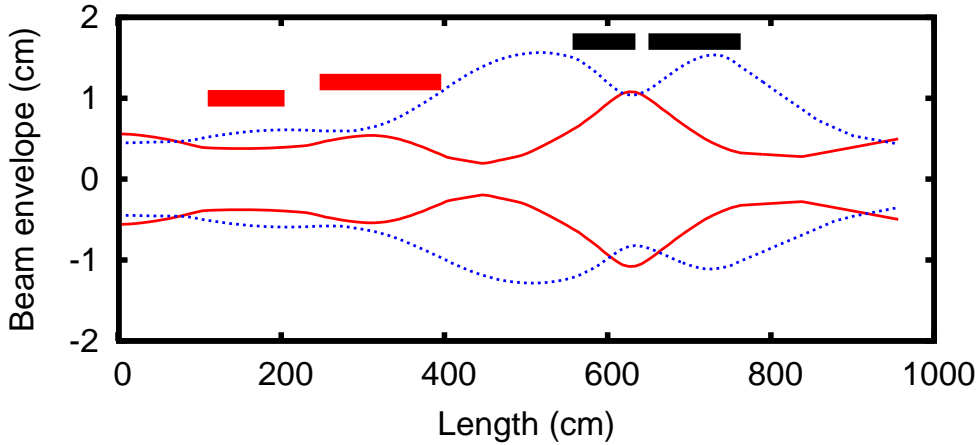}}
\end{center}
\caption{Radial (blue lines) and axial (red lines) beam envelopes along the extraction trajectory, with a uniform energy spread of $\pm 0.2$\%. The positions of electrostatic deflectors (ED1 and ED2) and magnetic channels (MC1-2) are also shown in red and black, respectively.
\label{fig:ch3-6}}
\end{figure}

\begin{table}[hbt]
\centering
\caption{Parameters of the DAE$\delta$ALUS Injector Cyclotron}
\begin{tabular}{lrrllrrl}
\hline
$E_{max}$ & 	60 MeV/amu	&	$E_{inj}$ & 35 keV/amu \\
$R_{ext}$ &	1.99 m		&		$R_{inj}$ &55 mm  \\
$<B>$ @ $R_{ext}$ &1.16 T	 &	$<B>$ @ $R_{inj}$ &	0.97 T  \\
Sectors		& 4			&		Hill width	&	28 - 40 deg \\
Valley gap	& 1800 mm	& Pole gap	& 100 mm  \\
Outer Diameter & 6.2 m	 & Full height & 2.7 m  \\
Cavities	& 4					& Cavity type	& $\lambda/2$, double-gap  \\
Harmonic &	 6th		&			rf-frequency	& 49.2 MHz  \\
Acc. Voltage	& 70 - 250 kV	 & Power/cavity &	$<110$ kW  \\
$\Delta E$/turn	 &1.3 MeV	& Turns &107  \\
$\Delta R$ /turn @ $R_{ext}$	& $20$ mm	 & $\Delta R$/turn @ $R_{inj}$ & $>56$ mm  \\
Coil size & 200x250 mm$^2$ & Current density	 & 3.1 A/mm$^2$  \\
Iron weight & 450 tons	& Vacuum  & $< 10^{-7}$ mbar  \\
\hline
\end{tabular}
\label{tab:injtab}
\end{table}

\subsection{Extraction from the DIC} 
\begin{figure}[ht!]
\begin{center}
{\includegraphics[angle=-0, width=0.6\linewidth,trim=0.0cm 3.0cm 0.0cm 2cm]{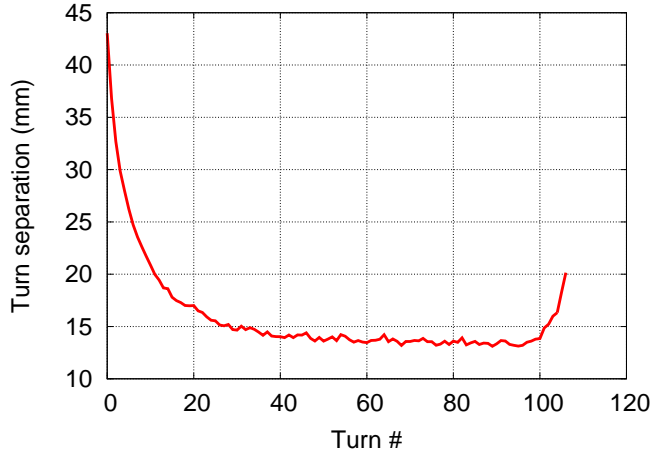}}
\end{center}
\caption{Turn separation of concentric orbits in the DIC. The radial gain is increased by about 7 mm  without exceeding a phase slip of $50^\circ$ 
\label{fig:ch4-7}}
\end{figure}

The turn separation of the concentric orbit is given as
\begin{equation}\label{eq:radgain}
\frac{dR}{dn} = R \frac{E_g } {E} \frac{\gamma}{\gamma+1}\frac{1}{\nu^2_r}
\end{equation} 
with $R$ denoting the radius, $E_g$ is the energy gain per turn, $E$ is the total energy of the
particles, $\gamma$ is the relativistic factor, and $\nu_r$ is the radial focusing frequency \cite{Joho}.
The development of the turn separation is shown in Fig.\ \ref{fig:ch4-7}; most importantly we get a turn separation of 20 mm at the extraction, which is enough for a clean extraction. 

\begin{figure}[ht!]
\begin{center}
{\includegraphics[width=0.6\linewidth]{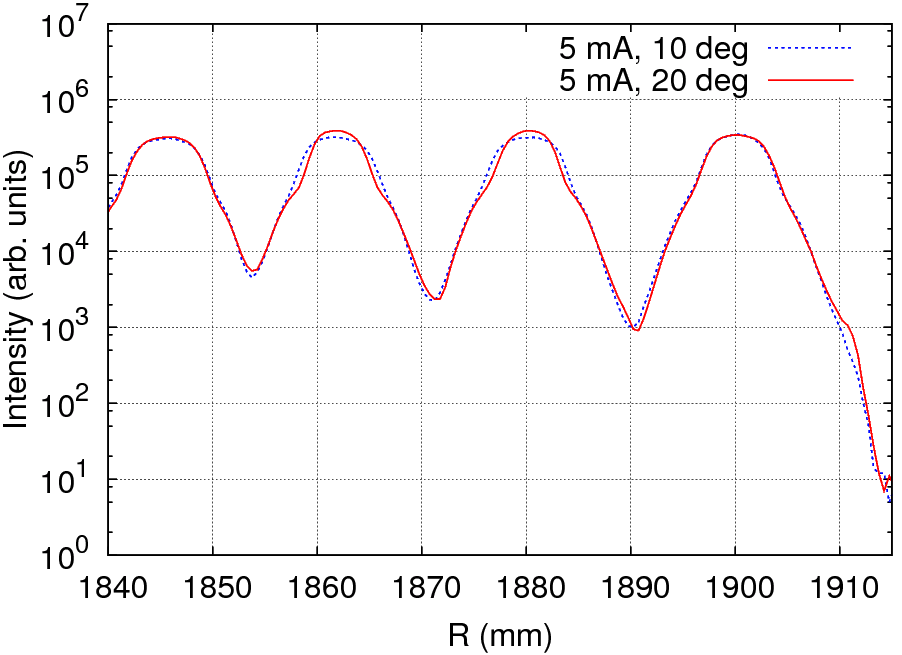}}
\end{center}
\caption{The radial profile of the last four turns at the center of the valley for a 5mA \htp\ beam with the initial phase width of $10^\circ$ and  $20^\circ$, respectively
\label{fig:ch4-8}}
\end{figure}

In Fig.\ \ref{fig:ch4-8} the radial profile of the last four turns are shown, at the center of the valley. In order to have enough statistics, we are using $10^6$ macro particles in
the full 3D simulation with space charge, similar to an earlier study \cite{PhysRevSTAB.14.054402}.
The phase acceptance of the central region is not yet known; hence, we assume practical values of $10^\circ$ and $20^\circ$.
If the electric channel is inserted at R=1.89 m with an effective septum thickness of 0.5 mm, 
then the extraction efficiency is better than 99.9998\% in the 5 mA case for both initial phases. 

The first electrostatic deflector ED1 is placed in a valley, inside of one of the rf-cavities. This solution has already been proven in other cyclotrons and offer the advantages of being able to place the ED in a region where the magnetic field is low and there is enough room in the vertical direction. In Fig.\ \ref{fig:ch3-6} the positions of ED1, ED2, and of the two additional passive magnetic channels that guide the beam out of the cyclotron yoke, are shown. The ED1 length and its gap are 28$^\circ$ and 20 mm respectively, while the applied electric field is 30 kV/cm. The ED2 length and its gap are 40$^\circ$ and 24 mm respectively, while the applied electric field is  25 kV/cm. The first magnetic channel (MC1), 10$^\circ$ long, is placed 36$^\circ$ after the end of ED2, at the entrance side of the following hill. The MC1 produces a shielding of the local magnetic field of about 2.5 kGauss. The MC2 is 26$^\circ$ long and its shielding field is 4.4 kGauss. At the entrance of the first magnetic channel the separation between the accelerated orbit and the extracted trajectory is about 80 mm; hence, the two beams will be well separated. The radial and axial beam envelopes of the beam along the extraction trajectory are  shown in Fig.\ \ref{fig:ch3-6}. The initial radial size is about 10 mm and represents the beam core. The beam envelopes also include the effects due to the uniform energy spread of  $\pm$ 0.2\%. The initial beam size was evaluated assuming a normalized beam emittance of 3.3 $\pi$ mm mrad. This value is about 30 times larger than the normalized beam emittance of the VIS.

\section{The DAE$\delta$ALUS superconducting ring cyclotron}
The DAE$\delta$ALUS Superconducting Ring Cyclotron is the main component in the accelerator chain. One of the big challenges again is beam loss, to be limited to a few hundred watts in total. This corresponds to relative beam losses in the lower $10^{-4}$ range. The choice of \htp\ for the accelerated molecule species results in weaker space-charge forces and thus a higher intensity limit. However, the charge-to-mass ratio of 1:2 calls for high magnetic fields, posing the second challenge in the DSRC design. 

The \htp\ beam will be injected along the cyclotron valley by four injector magnets and one electrostatic deflector and it will be accelerated with four single-gap cavities. The extraction is performed by insertion of a pyrolytic graphite stripper foil with thickness less than 2 mg/cm$^2$. Ionization through residual gas interaction of the beam contributes to the loss rate. A tentative layout of the DSRC is presented in the following sections. It includes the superconducting sector magnets, capable of generating sufficient field levels; the concept for the extraction foil in view of sufficient lifetime; and the overall design of the vacuum system that should provide pressure levels of $\approx 10^{-8}$\,mbar. 
The main purpose of the present magnet design is to demonstrate the feasibility of the physics parameters, but we are also taking into account that full optimization, particularly in coil design, may require substantial changes to this baseline design.

\subsection{The sector magnet for the DSRC}
\begin{figure}[ht!]
\begin{center}
{\includegraphics[angle=-0, width=0.4\linewidth]{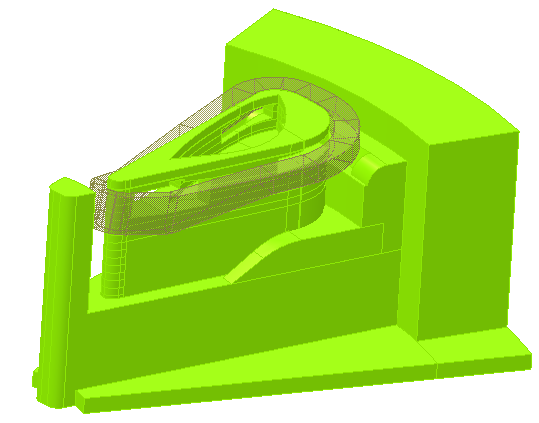}}
\caption{Drawing of the pole and the coil package
\label{fig:dsrc2}}
\end{center}
\end{figure}

In the DSRC the main magnet configuration consists of eight superconducting sector magnets. Compared with a previous study \cite{calabretta-2010,calabretta-2011-1,calabretta-2011-2}, the parameters of the DSRC magnetic sector have been 
significantly improved with respect to the shape and current densities of the superconducting coils to improve focusing properties and reduce forces on the structure. 
The iron of the hill nearest to the median plane now has a strongly modulated shape to achieve stronger vertical focusing. A hole in the center of the hill has been introduced to adjust the average field. The minimum clearance between the lower and upper cryostat surfaces is now at 75 mm, and the gap between the poles is at  80 mm in order to achieve sufficient vacuum conductance. Figure \ref{fig:dsrc2} shows the iron and conductor package of one sector, and Table\ \ref{tab:dsrcmag1} shows the main parameters of the coils.
\begin{table}[hbt]
\centering
\caption{The main parameters of the superconducting coils}
\begin{tabular}{lrlrl}
\hline
Coil size near center (width x height) & $15\times \text{48 cm}^2$ \\
Coil size outer region & $24 \times \text{30 cm}^2$ \\
I coil	& 3400 \text{A/cm}$^2$ \\
Max. Field & 6.18 T \\
\hline
\end{tabular}
\label{tab:dsrcmag1}
\end{table} 

\subsubsection{Focusing properties}

\begin{figure}[ht!]
\begin{center}
{\includegraphics[angle=-0, width=0.49\linewidth,trim=2.0cm 4.0cm 0.5cm 2cm]{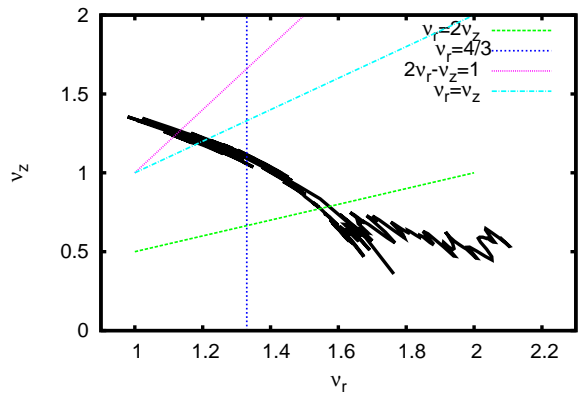}}
\end{center}
\caption{Tune diagram, $\nu_z$ vs. $\nu_r$ 
\label{fig:dsrc6}}
\end{figure}

The isochronous magnetic field error is $\pm$0.4\%, with sufficient focusing properties in both the radial and vertical planes. The $\nu_z$ is kept above 0.5 and the Walkinshaw resonance $\nu_r=2\nu_z$ is crossed quickly and only once at the beginning of the acceleration, as shown in Fig.\ \ref{fig:dsrc6}. The present baseline design  still poses several technical challenges, which in the future may require 
adjustments to the design as presented.

\begin{figure}[ht!]
\begin{center}
{\includegraphics[angle=-0, width=0.4\linewidth,trim=2.0cm 4.0cm 0.5cm 4cm]{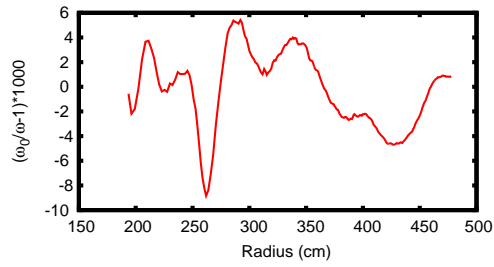}}
\end{center}
\caption{Relative difference between the ideal revolution frequency $\omega_0$ and the revolution frequency $\omega$ of the particles.
\label{fig:dsrc5}}
\end{figure}

Figure \ref{fig:dsrc5} shows the difference between the theoretical revolution frequency and the revolution frequency evaluated by the magnetic field map of the model presented here. The isochronism was evaluated by the orbit code Z4 \cite{genspe-1}.    
The radial and vertical focusing frequencies $\nu_z$ and $\nu_r$ are shown in Fig.\ \ref{fig:dsrc6}.
The fast oscillations are due to the grid size of the magnetic field map used to evaluate the model. 

\subsubsection{The analysis of magnetic forces}

\begin{figure}[ht!]
\begin{center}
{\includegraphics[angle=-0, width=0.98\linewidth]{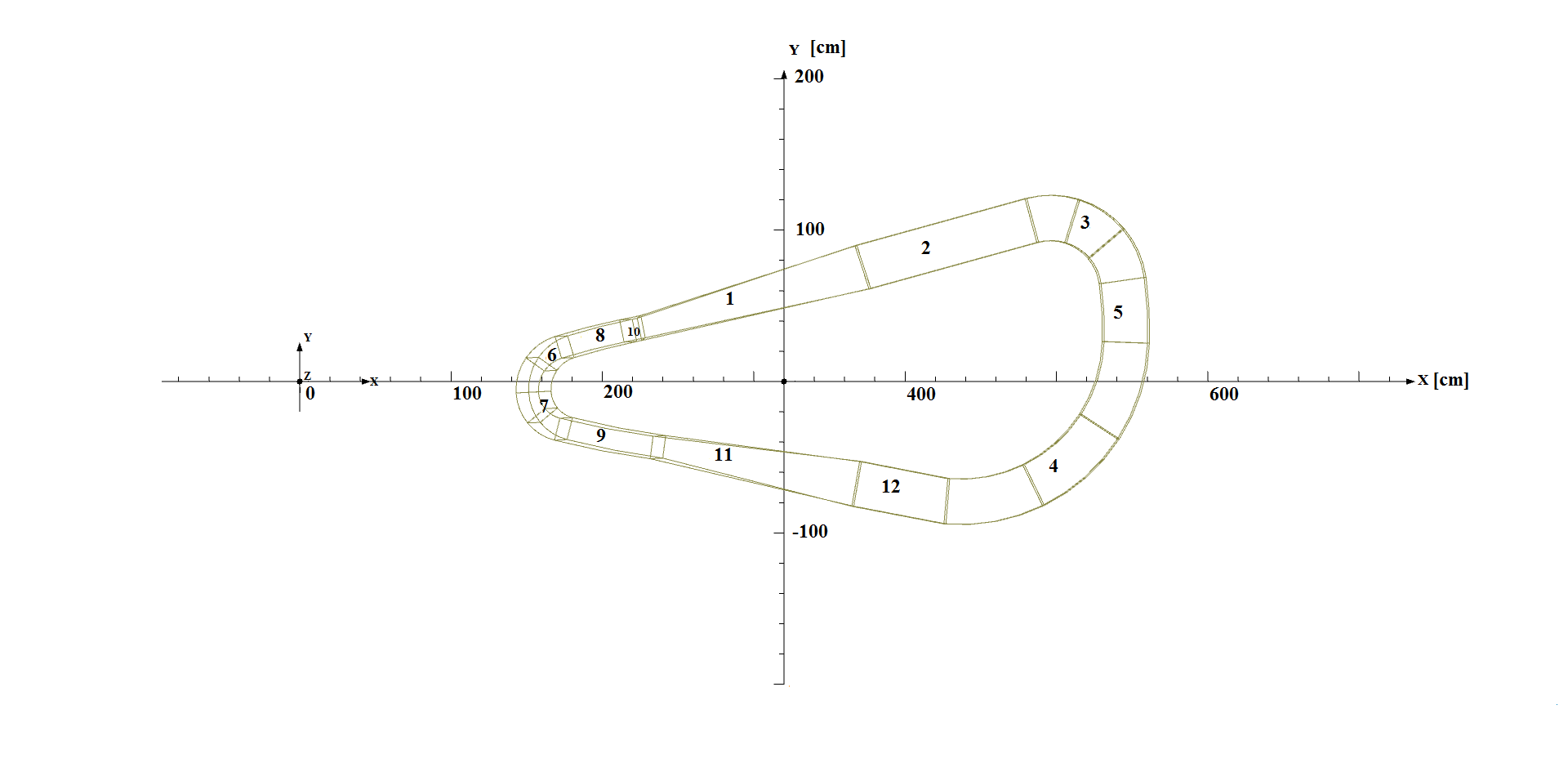}}
\end{center}
\caption{For the stress analysis, the superconducting coil is divided into 12 sections.\label{fig:dsrc3a}}
\end{figure}

A detailed evaluation of the magnetic forces resulting in structural forces is made by
using the code OPERA. The geometry and different sectors for the force evaluation are shown in Fig.\ \ref{fig:dsrc3a}.
The maximum field on the coil surfaces is lower than 6 T for all conductor segments, except on segments number 3 and 4. In these segments, the peak field reaches 6.18 T.
Strong expansion magnetic forces, which try to make the coil round, acting on segments 1, 2, 8, 9, 11, and 12.\ The strength of these forces, on the order of MN, is due to the high magnetic field and the large cross section of the coils. To compensate for these forces, different solutions have been evaluated. In the present design, the thickness of the LHe vessel varies. A thickness of 4 cm is chosen for the side near to the median plane and for the wall that is the nearest to the iron of the hill.  The other wall and the side farther from the median plane are 7 cm thick. To optimize the magnetic field configuration, the cryostat wall that faces the pole has to be made of iron.  The outer part that faces the valley and the median plane has to be stainless steel. The use of iron and stainless steel to build a cryostat has been already demonstrated.

An additional solution to the huge forces issue was suggested in \cite{okuno-2008,okuno-2004}. Opposite segments are connected by two stainless steel plates \cite{calabretta-2011-2}.  As the forces on the opposite segments of the coil have roughly the same magnitude but opposite directions, the residual force can be nearly canceled. 

The thicknesses of the LHe vessel and of the connecting plates are chosen to maintain the stress values below 6 dN/mm$^2$. The region between the LHe vessel and the cryostat walls is in a vacuum to guarantee the cryogenic insulation. The distance between the inner wall of the cryostat and the outer wall of the LHe vessel is fixed at 40 mm and it is sufficient to accommodate more than 30 layers of  aluminized insulators. 

To minimize the thermal load on the LHe vessel an intermediate shield, cooled to 77 K, is installed in the gap between the cryostat and the LHe vessel. Enough room is included above and below the cryostat for an elliptical cooling tube for liquid nitrogen. 

The attractive magnetic force between two symmetrical coils, with are symmetrical with respect to the median plane, is large. It is on the same order as in the RIKEN SRC design. Also, in agreement with RIKEN, the present design assumes the coil segments 6 and 7 at the inner radii of the upper coil and the coil segments 3, 4, and 5 at the outer radii of the upper coil are connected with their corresponding segments of the lower coil by stainless steel plates, which are large enough to counterbalance this attractive force. Note that the parts of the cryostat corresponding to these segments are more solid than the ones corresponding to the other segments, which are sustained only by the structure of the LHe vessel. 

In Table\ \ref{magnet1} we compare the main parameters of the DSRC magnet design with the parameters of the existing RIKEN SRC. 
Many parameters are similar; hence, they are creditable in lieu of a technical realization.
\begin{table}[ht!]
\centering
\caption{Comparison of the main parameters for DSRC with those for RIKEN-SRC}
\begin{tabular}{lrrl}
\hline
\textbf{Basic Parameters} & \textbf{DSRC } & \textbf{RIKEN-SRC} & \textbf{Unit} \\
\hline
Maximum field on the hill& 6.05 &3.8 &T\\
Maximum field on the coil& 6.18 &4.2 &T\\
Stored Energy & 280 & 235 & MJ \\
Coil size & 30$\times$ 24 or 15$\times$ 48 & 21$\times$ 28 & cm$^2$ \\
Coil Circumference & 9.8 & 10.86 & m \\
Magnetomotive force& 4.9 & 4 & MAtot/sector \\
Current density & 34 & 34 & A/mm$^2$ \\
Height& 5.6 & 6.0 & m \\
Length& 6.9 & 7.2 & m \\
Weight& $\leqq$450 & 800 & ton \\
Additional magnetic shield& 0 & 3000 & ton/total \\
\hline
\textbf{Magnetic Forces} &&&\\
\hline
Expansion& 1.87 or 1.8 &2.6 &MN/m\\
Vertical& 3.7& 3.3 &MN \\
Radial shifting& 2.7& 0.36 &MN \\
Azimuthal shifting & 0.2 & 0 &MN \\
\hline
\textbf{Main Coil} &&&\\
\hline
Operational current& 5000&5000 &A\\
Layer $\times$ turn & 31$\times$16& 22$\times$18 & \\
Cooling& Bath cooling & Bath cooling & \\
Maddock Stabilized Current& 6345& 6665 &A \\
\hline
\textbf{Other Components} &&&\\
\hline
SC trim& no&4 &sets\\
NC trim $\times$ turn & no & 22 &pairs \\
Stray field in the SRC valley region& 0.7 & 0.04 & T \\
Gap for thermal insulation& 40 & 90@min. &mm \\
Extraction method& Stripper foil&Electrostatic channel& \\
\hline
\end{tabular}
\label{magnet1}
\end{table}

\subsection{Radial injection into the DSRC}
 We assume the \htp\ beam has an energy of 60 MeV/amu, an rms energy spread of 0.2\%, and a normalized emittance of $10.0 \times \pi\text{ mm } \text{mrad}$ in order to match a beam-size  similar to the one predicted by \opal. The main constraints of the injection scheme are the following:
\begin{itemize}
\item the axial beam envelope must be smaller than 5 cm (full width) in the region where the beam crosses the accelerating gap
\item the beam has to match the equilibrium orbit at the matching point, which is chosen at $\theta=180^\circ$ in the middle of the hill
\item in order to bend the trajectory we plan to mainly use a passive magnetic channel. Normal conducting coils will be used to tune the fields.
\end{itemize}
\begin{figure}[ht!]
\begin{center}
{\includegraphics[angle=-0, width=0.8\linewidth,trim=2.0cm 6.0cm 0.5cm 4cm]{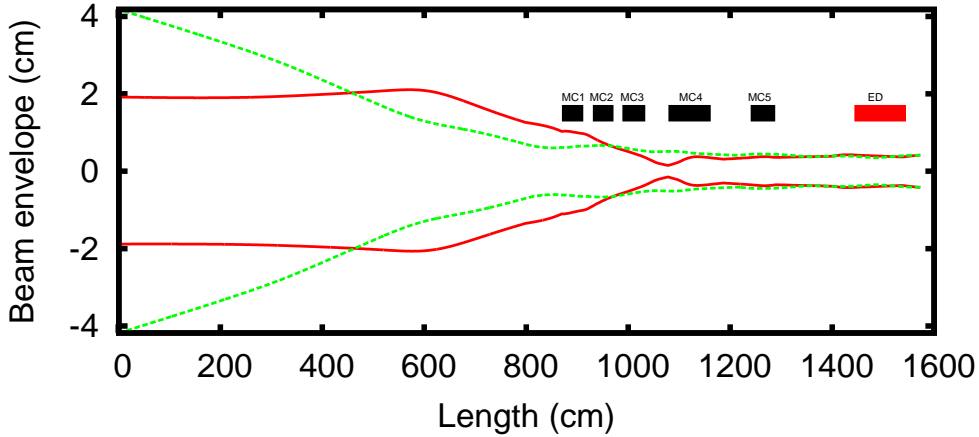}}
\end{center}

\caption{Vertical (green) and radial (red) envelope along the injection trajectory, computed using the OPERA code until ED. From ED to the electrostatic deflector the core Z4 \cite{genspe-1}  is used.
\label{fig:dic1}}
\end{figure}
To achieve these constraints, we track eigen-ellipses backwards from the matching point. The results of Z4 \cite{genspe-1} simulations are shown in Fig.\ \ref{fig:dic1}. Here the injection trajectories are shown together with the main components, the five passive magnetic channels (MC1-MC5) and the electrostatic deflector (ED). The trajectory crosses a field-valley with a rf-cavity installed however; the effect is a slight radial steering of the trajectory that can be compensated for by an external steering magnet.  

MC2 is placed just inside a hole in the inner return yoke of the sector magnet and has to increase the shielding effect of the yoke by about $-2$ kGauss. MC3 stays in the inner fringing field and increases the local stray field to $11$ kGauss. The most critical device is MC4, which has to produce a shielding effect of $-13.8$ kGauss. This large value can be achieved by using a passive shielding magnet and an additional normal conducting coil to produce a tuning field in the range of about $-1$ kGauss.
The separation between the equilibrium orbit and the injection trajectory at the position of the magnetic channels is 4 cm at the minimum. In particular, the orbit separation exceeds 20 cm at MC3 and MC4, the channels that need higher magnetic fields. 

The separation between the injected beam and the first accelerated orbit at the position of the ED is about 2 cm. This distance could be increased up to 3 cm by injecting the beam slightly off-center. This first harmonic should produce a beam oscillation which increases the separation up to 3 cm at the position of the ED. The deflecting voltage of the electrostatic element is at a conservative value of 50 kV.

\subsection{Stripper extraction} \label{sec:extstripp}
The stripping method utilizes a thin pyrolytic graphite foil, highly oriented, with a thickness of about 2 mg/cm$^2$. By interaction with the foil, the molecule dissociates into two protons which follow an orbit with different curvature due to their charge-to-mass ratio of one. The stripping extraction has an efficiency close to 
100\%, even if the beam orbits at the extraction radius are not well separated. A disadvantage is the relatively high energy spread of the extracted beam due to overlap of several turns on the foil. According to OPAL simulations, an energy spread of 1.2\% is expected. 

\begin{figure}[ht!]
\begin{center}
{\includegraphics[angle=-0, width=0.8\linewidth]{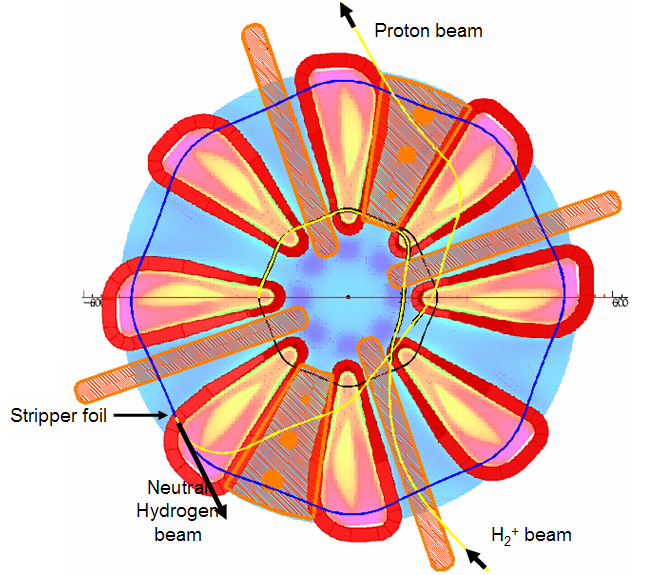}}
\end{center}
\caption{Injection and extraction scheme of the DSRC. The extraction trajectory for the proton beam with an energy of 800 MeV and the injection trajectory of \htp\ with an energy of 60 MeV/amu are shown as yellow trajectories. The stripper is placed at an angular position of $215^\circ$. The magnetic channels along the injection trajectory and the extraction trajectory are indicated.
\label{fig:dsrc8}}
\end{figure}

\begin{figure}[ht!]
\begin{center}
{\includegraphics[angle=-0, width=0.8\linewidth]{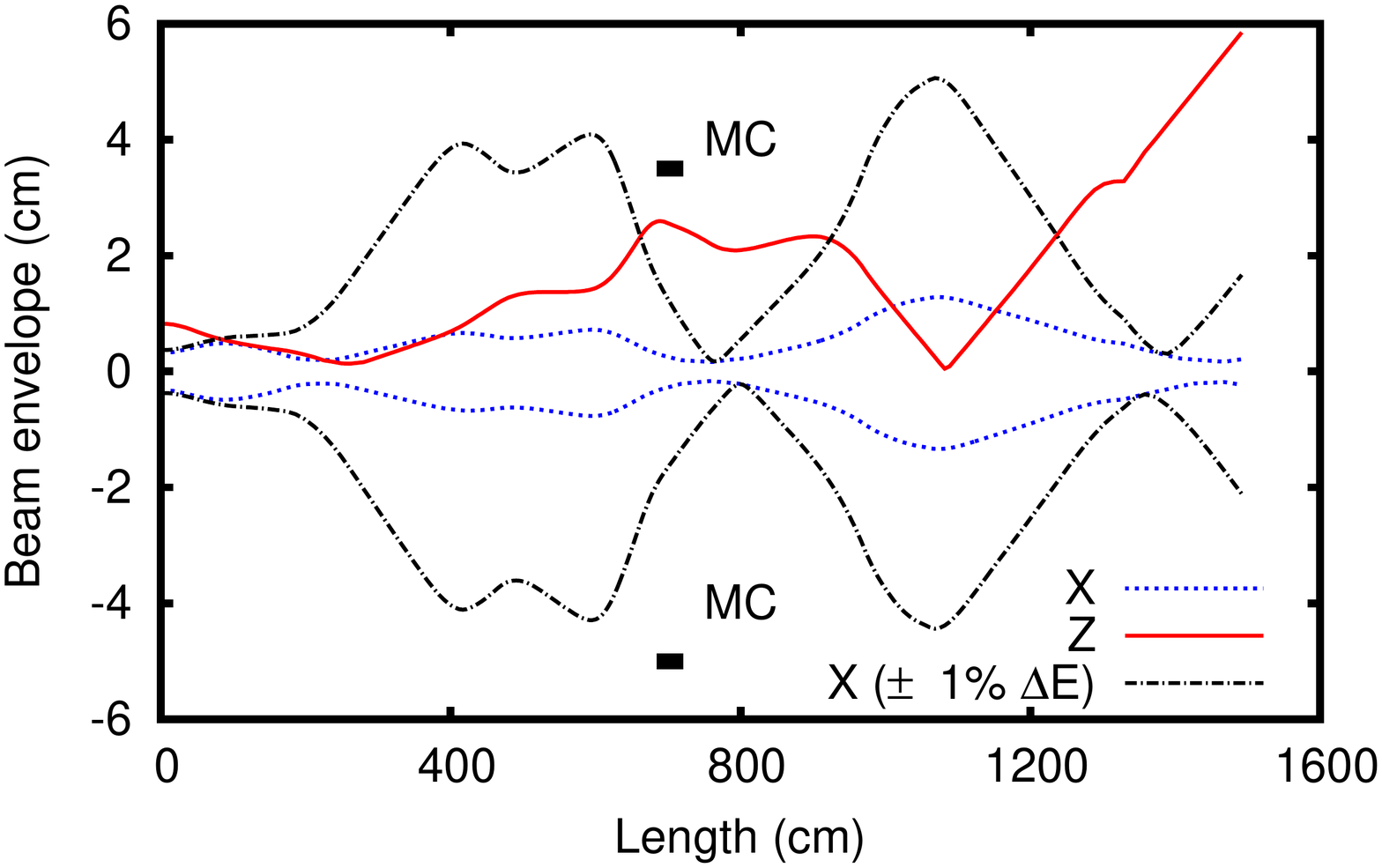}} 
\end{center}
\caption{The radial (blue lines) and axial (red line) envelopes for the proton beam with zero energy spread vs. distance from the stripper position along the extraction trajectory are shown. The black line indicates the radial beam envelope for an energy spread of $\pm$ 1\%. The position of the steering magnet channel (MC) is indicated as black boxes. 
\label{fig:dsrc9}}
\end{figure}

Figure \ref{fig:dsrc8} shows the extraction trajectory. The stripper foil is placed on the entrance edge of the hill, allowing us to safely extract a small fraction ($<10^{-4}$) of neutral hydrogen particles that are produced during the stripping process.
The beam envelope vs. beam trajectory length is shown in Fig.\ \ref{fig:dsrc9}. This figure shows the radial and axial beam envelope for the 800 MeV proton beam with and without $\pm$1.0\% energy spread. The axial beam envelope does not change significantly due to
the energy spread. The maximum radial beam envelope due to the energy spread is ~+5 cm/ -4.5 cm. At the exit of the cyclotron, the radial width of the beam envelope is less than 3 cm.  
The trajectories originate on the stripper foil placed at $R=488.3$ cm and at the azimuth angle of $215^\circ$. The beam envelopes in the radial and axial plane were obtained from the eigen-ellipses evaluation. The normalized emittance is assumed to be $10 \times \pi\text{ mm } \text{mrad}$ (corresponding to 100\% of the beam), and again roughly 100 times larger than the ion source emittance.

At the position where the trajectory is nearest to the center of the cyclotron, it is possible to install a magnetic channel to steer the trajectory and to provide axial focusing in order to maintain the vertical beam envelope below 4 cm along the trajectory inside the vacuum chamber. The angular extension of the magnetic channel is $10 ^\circ$, inside an empty valley. The bias field of the magnetic channel is 0.1 T and its radial gradient is 0.1 T/cm. The maximum size of the beam inside this magnetic channel is about 8 cm.
The stripper foil azimuth position was chosen to stay in the region where the magnetic field is in the range $0.2$ to $0.4$ T. The bending radius of the 
stripped electron will be $\sim$ 8 to 9 mm; hence, these electrons can be stopped easily on a copper shield.

\subsection{Stripper foil considerations}
At the outer radius of the DSRC, \htp\ ions are dissociated into two protons by a carbon stripping foil and are led into the extraction channel that snakes through the central region of the DSRC.  Various considerations are important: a) foil type, shape, thickness, and mounting technique; b) foil lifetime; and c) stripping efficiency.

\subsubsection{Foil type}
Substantial experience exists with strippers for high-current beams for stripping of \hmi\ from machines such as TRIUMF, Spallation Neutron Source (SNS), and commercial (lower-energy) isotope-producing cyclotrons. TRIUMF utilizes 2 mg/cm$^2$ highly oriented pyrolytic graphite foils for their 520 MeV beam, supported on an inverted L-shaped frame which is in turn also supported on two edges so as to minimize mechanical constraints from foil changes due to thermal cycling and aging.  In addition, clamping the carbon stripper, a tantalum frame that is lined with thin, wrinkled copper foil helps reduce mechanical stresses \cite{bylinski-1}.  SNS strips 1 GeV ions using 260 $\mu$g/cm$^2$ polycrystalline diamond foils supported only from the top edge, thinner foils being required to minimize scattering due to multiple passes of the stripped beam during stacking in the accumulator ring \cite{plum-2008}. Commercial isotope cyclotrons for 30-70 MeV \hmi\ use 200-400 $\mu$g/cm$^2$ pyrolytic graphite supported on a C frame.

\subsubsection{Foil temperature \& lifetimes}
The foil lifetime is highly dependent on the average and instantaneous power deposition from the beam, and on the trajectory of electrons stripped from the primary ions.  
Furthermore, the pulse structure can have a substantial impact on the peak foil temperature, as radiative cooling time constants play an important role.
Ultimately, the peak temperature in the foil determines its lifetime, and temperatures above 2500 K lead to rapid failures due to sublimation. Generally, the lifetime is inversely  proportional to the beam intensity, given that the foil temperature is low. In fact, foil lifetimes are affected by thermionic emission as well, which becomes appreciable at temperatures even below 2000 K (where graphitization starts); hence, controlling peak temperatures is of great importance. 

\begin{figure}[ht!]
\begin{center}
{\includegraphics[angle=-0, width=0.49\linewidth]{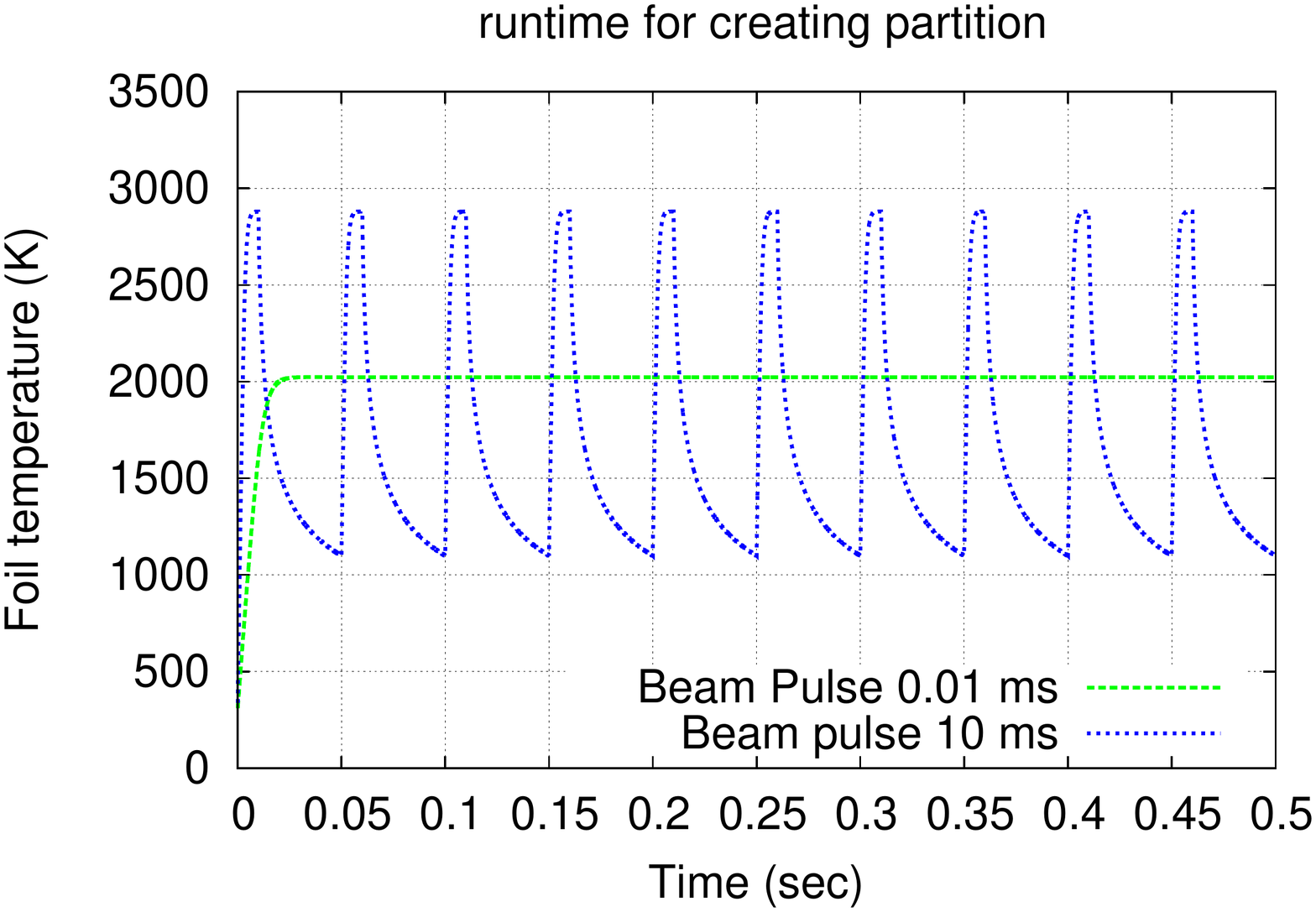}}
{\includegraphics[angle=-0, width=0.49\linewidth]{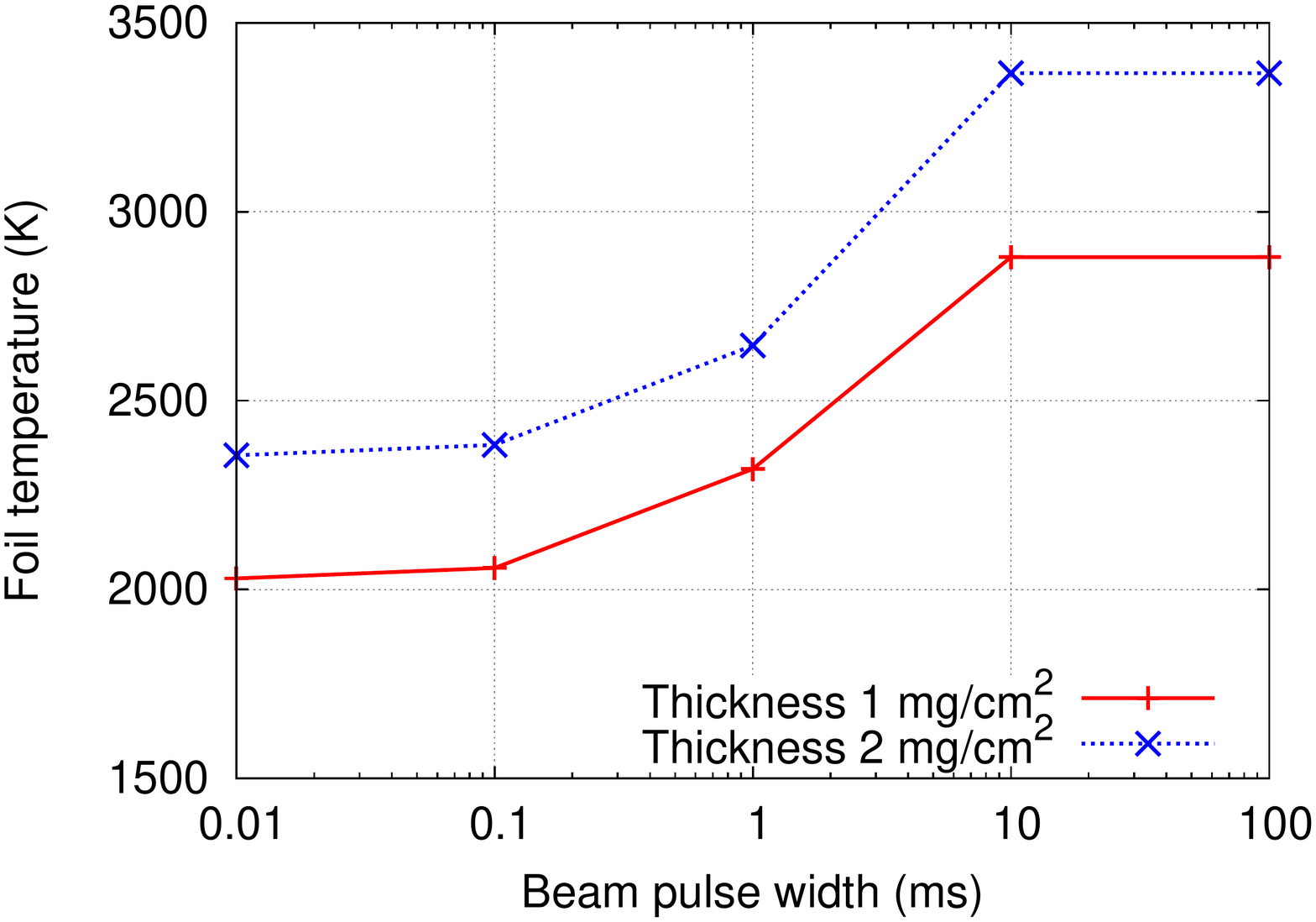}}
\end{center}
\caption{Thermal cycling in the foil vs. pulse width:  shorter pulses lead to much more regular foil heating (left figure). Maximum foil temperature vs. pulse width. The knee is at about 1 ms pulse width, at a rate of  200 Hz (right figure). 
\label{fig:foil-1} }
\end{figure}

Instantaneous power is assumed to be a factor of 5 over average power operating at a duty factor of approximately 20\%.\  The foil temperature $\theta_{foil}$ was calculated by integration of the following equation:
\begin{equation*}
\frac{d \theta}{dt} = \frac{1}{\tau \rho c} \times (\omega -2\sigma \epsilon \theta^{4}),\\
\end{equation*}
\begin{equation*}
\theta_{foil}=\theta + \theta_{circ}
\end{equation*}
where all the variables and constants are defined in Table\ \ref{tab:foil1}.
\begin{table}[ht!]
   \centering
   \caption{Estimation of the foil temperature for a foil thickness of $2~mg/cm^2$ and $1~mg/cm^2$}
   \begin{tabular}{lrrl}
       \hline
       \textbf{Quantity} & \textbf{ foil thickness $2~mg/cm^2$} &  \textbf{foil thickness $1~mg/cm^2$} & \textbf{Unit}\\
       \hline
           Energy loss&0.0044&0.0022&MeV\\
           Peak current \htp&5&5&mA\\
           Duty cycle&20&20&\%\\
           Pulse width&0.1 $\sim$ 1000&0.1 $\sim$ 1000&ms\\
           Beam Area (4 x rms)&10&10&mm$^2$\\
           Foil thickness $\times$ density $\tau \rho$&2&1&mg/cm$^2$\\
           Specific Heat $c$&8.5/12&8.5/12&J/g K\\
           Max. power density during beam-on $\omega$&10&5&W/mm$^2$\\
           Emissivity $\epsilon$&1&1&\\
           T circumference $\theta_{circ}$&303&303&K\\
       \hline
   \end{tabular}
   \label{tab:foil1}
\end{table}
The average power density during beam on can be estimated to be about 1~W/mm$^2$, assuming that beam profile is uniform; however, a $\omega$ of  3~W/mm$^2$ 
was used from the beam profile at the extraction foil (Fig.\ \ref{fig:ch2-2}). The left graph in Fig.\ \ref{fig:foil-1} shows a temperature evolution in the cases of pulse widths of 0.01~ms and 1~ms. In the case of 0.01~ms, the temperature is determined by the average beam power. In the case of 10~ms, we observe large temperature fluctuations with a peak temperature at 2500 K. 
The right graph in Fig.\ \ref{fig:foil-1} shows the foil maximum temperature as a function of pulse width, from $0.01 \dots 10$ ms. From the graph it is obvious that the pulse width should be less than 1~ms to keep the foil temperature at acceptable levels.

Energy deposition on the foil can be severely affected by the orbits of convoy electrons.  These stripped electrons carry an energy of about 440 keV each, so contribute an average power of 272 watts per megawatt of protons.  If the foil is placed in a strong field, the electrons spiral around and deposit this energy into the foil.  To control this, the plan is to place the foil in the fringe-field of a sector magnet, where the strength is about 0.2 T, so the electron orbital radius is about 1 cm.  As the electrons are bent towards the outside, a cooled electron catcher can be placed to intercept them.

The performance experience of SNS can be taken as an indication that acceptable foil lifetimes are well-within current technology.  The SNS runs at a 6\% duty factor, with 1 ms beam on and 16 ms beam off, and an average power of 1 MW (17 MW peak power).  The Chemical Vapor Deposited (CVD) polycrystalline foils have operational lifetimes of several months \cite{plum-2008,plum-1}. Possible reduction of foil thickness to account for what is expected to be more efficient stripping of the one electron from \hmi\ as compared to the two from  \htp\ (one which is rather tightly bound), could lead to a reduction in energy loss (power deposition) in the foil, and even longer lifetimes.

\subsubsection{Stripping efficiency}
Incomplete stripping can lead to beam loss at high energy, and if not mitigated, will be a major contribution to the power-loss budget of the DSRC. Incomplete stripping occurs when the primary beam misses the foil, or if ions emerge either as unstripped \htp\ or \hze. The foil size will be a few cm$^2$, and will cover the full vertical height of the beam. In the radial dimension, the foil will cover the last several turns; hence, the probability of an ion not intercepting the foil is vanishingly small.  Furthermore, if an ion misses the foil (on its inner edge), it will intercept the foil on its subsequent turn.  

A foil thickness of several hundred $\mu g/cm^2$ is sufficient to break up all molecular ions passing through it.  Consider now the extremely unlikely event of an intact (not stripped) ion that emerges from the foil; in that case the energy loss of this ion is about $\frac{1}{2}$ MeV per mg/cm$^2$ of foil thickness, which is substantially less than the energy gain per turn, so the ion will intercept the foil again on its subsequent turn.  

Neutral hydrogen is a significant problem in stripping  \hmi; at SNS they observe about 3\% of their beam emerging from the stripper as \hze. A non-negligible fraction of these neutrals are in weakly bound, long-lived (metastable) Rydberg states. The fraction of neutrals is expected to be substantially lower for \htp, as there is only one electron to remove, whose (ground-state) binding energy is 2.7 eV; while the \hmi ion has two electrons. The most loosely bound at 0.7 eV is easy to strip, but the second electron, with a binding energy of 13.6 eV may be more difficult to cleanly remove.

Mitigation of the neutral components is relatively straightforward; ground-state neutrals will follow a straight-line trajectory from the stripper, and a dump can be placed at the point where these atoms exit the magnet yoke.
These ground-state neutrals will safely traverse even the strongest magnetic field region, because the binding energy is such that Lorentz stripping of these atoms is essentially impossible.  However, the weakly bound metastable neutrals will undergo Lorentz stripping when traversing any appreciable magnetic field, resulting in a  contribution to beam halo in the extraction channel.  However, the acceptance of the DSRC extraction channel is large, and simulations have indicated that all such halo particles will be extracted cleanly, as long as the magnetic field stripping these atoms is sufficiently close to the stripper foil. 

\subsection{Vacuum system}
Due to  interactions with the residual gas, ions can lose their orbital electron as they travel along the acceleration path. The fraction of beam particles which survives acceleration is given by \cite{Betz1972}
\begin{equation}
T = N/N_0 = \exp{(-3.35~10^{16} \int \sigma_L(E) P dL)},
\end{equation}
\begin{equation}
\sigma_L(E) \approx 4\pi a_0^2 (v_0/v)^2 (Z_t^2 + Z_t)/Z_i
\end{equation}
where $P$ is the pressure (torr) ($3.35 \times 10^{16}$ is the number of molecules/cm$^3$ in one Torr), $L$ is the path length in cm, and $\sigma_L(E)$ is the cross section of electron loss. The ion velocity is denoted with $v$, $v_0$ and $a_0$ are the characteristic Bohr velocity and radius and, $Z_t$ \& $Z_i$ are the atomic number of the residual gas and of the incident ion respectively. While this formula is in good agreement with experimental data, its accuracy is not demonstrated for energies higher than 100 MeV/amu

Assuming that the distribution of vibrational states of \htp\ is sufficiently cold that the Lorentz-force induced dissociation is negligible, scattering of beam ions on the residual background gas will be the primary beam loss mechanism during the acceleration process. Therefore, keeping the total beam power lost in the DSRC to 200 W or less requires that the vacuum chamber remains at ultra-high vacuum conditions ($\leq$ 10 nTorr) despite the presence of a MW-class beam.

From Table 10.2 of \cite{prelimDaed}, the electrical current lost in the DSRC due to beam-gas collisions is
\begin{equation}
I_{ions} = \frac{I_{ave}}{2 \text{ mA}} \frac{P}{10 \text{ ntorr}} ~ 1.25 ~\mu A,
\end{equation}
where $I_{ave}$ is the average current in a macro pulse and $P$ is the gas pressure in the vacuum chamber. Converting this to protons lost per
second we get the following scaling relation,
\begin{equation}
\frac{dN_{beam}}{dt}=\frac{I_{ave}}{2 \text{ mA}} \frac{P}{10 \text{ ntorr}} ~ 7.25 \times 10^{12} \text{ ions/s}.
\end{equation}
For the nominal parameters, the energy deposition is $7.25 \times 10^{12} \text{ ions/s} \times 8~10^8 (eV) \times 1.6~10^{-19} = 560 \text{ W}$
which is too high by a factor of two. However this value is computed from the peak current derated by the duty factor. With respect to the pumping requirements which is probably pessimistic since the diffusion times for residual gas are relatively high -- thus the pressure peak smears out. It is also
evident, that with respect to the peak current the gas load is dominated by thermal outgassing. As the number of protons hitting the walls determines the activation of the chamber, as well as the dynamic gas load, a prudent choice may be to reduce the baseline pressure to 5 nTorr.

A critical engineering parameter in the design of the vacuum system is the effective ion-induced desorption coefficient, defined as 
\begin{equation}
\eta_{eff} = \frac{\text{number of molecules desorbed}}{\text{number of incident ions}}.
\end{equation}
The incident particles that produce the dynamic gas load, $Q_{dyn}$, can come both from the beam itself and from the fragments of residual gas that has been ionized by the beam and accelerated by the electromagnetic fields of the beam into the walls. 
If the vacuum system is free of leaks and if all seals are metal-to-metal except for the sliding seals between the rf-cavities and the beam chamber, the thermal out-gassing of the vacuum chamber walls, $Q_0$ will provide the only static gas load in the absence of beam. For an unbaked vacuum chamber, $Q_0$ may be as large as $2\times10^{-10}$ Torr-l/s/cm$^2$.  
The four rf-cavities introduce an additional area of $\sim 1.5\times10^6$ cm$^2$. As the cavities are run well above room temperature (assume T = 70 C), the out-gassing in these cavities may be as high as $\sim 20$ times larger than in the vacuum chamber. Consequently the total static load is $3.2 \times10^{-4}$ Torr-l/s.
The dimensionless ion-induced desorption coefficient, $\eta_{eff}$, depends strongly on the angle of incidence of the ion and on the ion energy. Relevant measurements of the ion-induced desorption by protons have been reported by the CERN vacuum group. 
\cite{mahner-08,cern-99}. The value is $\sim$ 10 for the gases mostly commonly found in ultra-high vacuum systems. To be very conservative one might double these values of $\eta_{eff}$ because the desorption coefficient will be larger before the walls are scrubbed by impacts from lost beam particles.

Therefore, the dynamic gas load\footnote{$1 \text{ molecule/s} = 2.83\times 10^{-20} \text{ torr-l}\text{/s}$} in the system is
\begin{equation}
Q_{dyn} = \eta_{eff} [ I_{ave} / \text{2 mA}] \times [ P / \text{ 10 ntorr} ] \times 4.1 \times 10^{-7} \text{ torr-l}\text{/s},
\end{equation}
and consequently, the total gas load reads: $Q = Q_{dyn} + Q_0$.
The pumping speed, $S$, required to maintain the cyclotron at a pressure, $P$, in the presence of the gas load due to ion-induced desorption is
\begin{equation}
S = \frac{Q}{P} =  \eta_{eff} \{[ I_{ave} / \text{2 mA}] \times [ P / \text{ 10 ntorr} ] \times 4.1 \times 10^{-7} + Q_0\} \times 10^8 / \text{ [P / 10 ntorr]}\text{[l/s]}  .
\end{equation}

With proper preparation of the chamber, we expect the value of $ \eta_{eff}$ to be $\approx$ 20 for \htp.
Assuming that for 1 to 10 ms pulses we must use the peak current in place of $I_{ave}$,
\begin{equation}
S = [I_{peak} / \text{2 mA}] \times 8.2~10^{2} + 32000  \text{ [1/s P / 10 ntorr]} \approx 4 \times 10^{4}  \text{ [l/s]}
\end{equation}
a number that is well within the capacity of sorption pumps.  

One attractive choice, which is effective at pumping hydrogen, would seem to be a cryopanel operating at 4.2 K. Measurements of a porous carbon panel at TRIUMF [4] achieve a pumping rate of 1.2 l s$^{-1}$ cm$^{-2}$; a CERN system using a silver coated surface does much better, delivering  9.2  l s$^{-1}$ cm$^{-2}$. As the coverage of hydrogen builds up on the cold surface, the equilibrium vapor pressure of $\text{H}_2$ will increase well beyond 10 nTorr. Therefore maintaining the pressure at 10 nTorr would require running the cryopanels in each sector magnet with an area of 1000 cm$^2$ each at $\sim$ 3.5 K.  An alternative used successfully at RIKEN is using high capacity cryopumps.  These have the advantage of being placed in areas of maximum gas-load, thereby minimizing the reduction of effective pumping speed due to limited conductance such as in the rf-cavities. Using eighteen pumps each of $\sim 10^4$ l/s, the RIKEN-SRC,  which has a larger vacuum chamber surface area than the DAE$\delta$ALUS design, obtains a base pressure of $< 10$ nTorr.  The pumps are regenerated every four to six months.  

An advantage of using the RIKEN approach is that high-capacity pumps can be placed directly at the top and bottom of the DAE$\delta$ALUS rf-cavities, which will be similar to the PSI design. As the rf-cavities constitute the largest fraction of the surface area in the vacuum system, this placement of pumps will be highly effective.

One region of concern is the area near the carbon stripping foils for extraction. If one loses 0.02\% of the beam in this region  similar to the losses near the septum in the PSI machine  the resulting dynamic load would be $Q_{dyn}  \sim 2.3 \times 10^{-6}$ Torr-l/s.

\subsection{Space-charge-effects}
\label{sec:SpCh}

For DSRC cyclotrons, single bunch space-charge effects are not the only contribution.   
The radial turn separation at injection is 27 mm and gradually reduces to about  3 mm at the extraction.  
In consequence, the beam size at the outer radii can be larger than the radial separation, and radially neighboring bunches will partially overlap. 
  
This simulation starts with an initial centered beam of 60 MeV/amu at injection. 
In order to evaluate the influence of space charge, simulations are done for the beam currents of 1 and 5 mA.
The initial transverse rms emittance $\epsilon$=0.9 $\pi$ mm-mrad (normalized) and the initial phase width and energy spread are set to a practical value of $10^{\circ}$ and 0.6\%, derived from the PSI operational experience.

The simulation result shows the vertical beam extension induced by space charge and neighboring bunch effects is small. 
For the 5 mA current, the beam halos extends vertically to 60 mm (full width). 
Considering the hill gap of the DSRC at 80 mm (full width), the beam is well separated from the magnet sector and vacuum chamber. 
There  remains enough space for the vertical beam envelop fluctuations caused by a possible mismatch of the injected beam.

\begin{figure}[ht!]
\begin{center}
{\includegraphics[width=0.6\linewidth,trim=0.0cm 3.0cm 0.0cm 2.0cm]{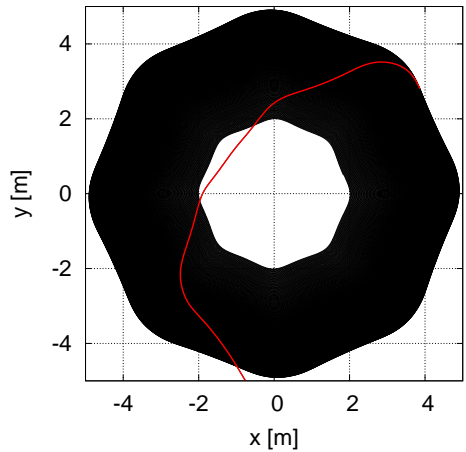}}
\caption{The \htp acceleration trajectory (black) and proton extraction trajectory (red)
\label{fig:ch2-3}}
\end{center}
\end{figure}

\begin{figure}[ht!]
\begin{center}
{\includegraphics[width=0.45\linewidth,trim=2.0cm 3.0cm 0.5cm 2cm]{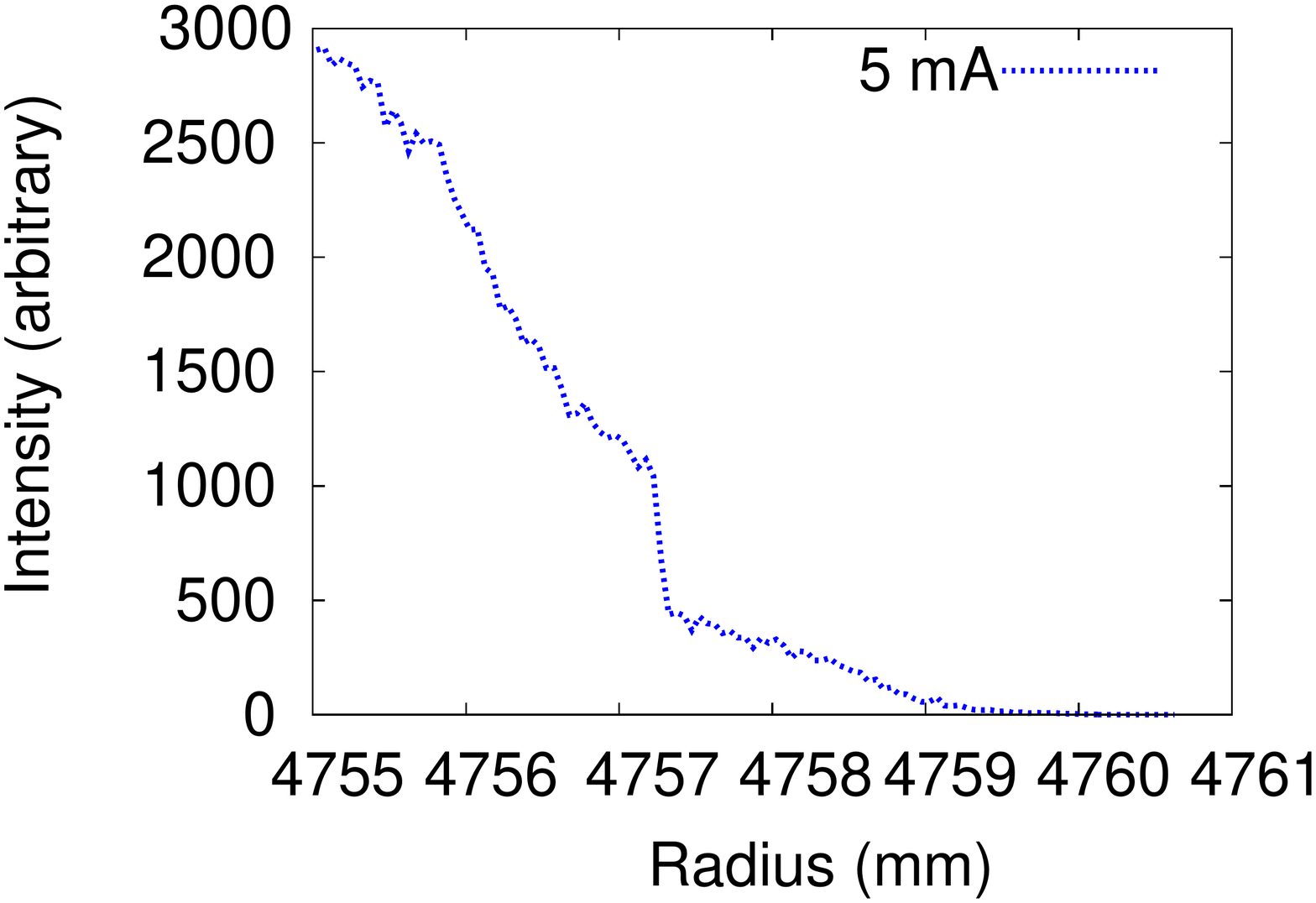}}
{\includegraphics[width=0.45\linewidth,trim=2.0cm 3.0cm 0.5cm 2cm]{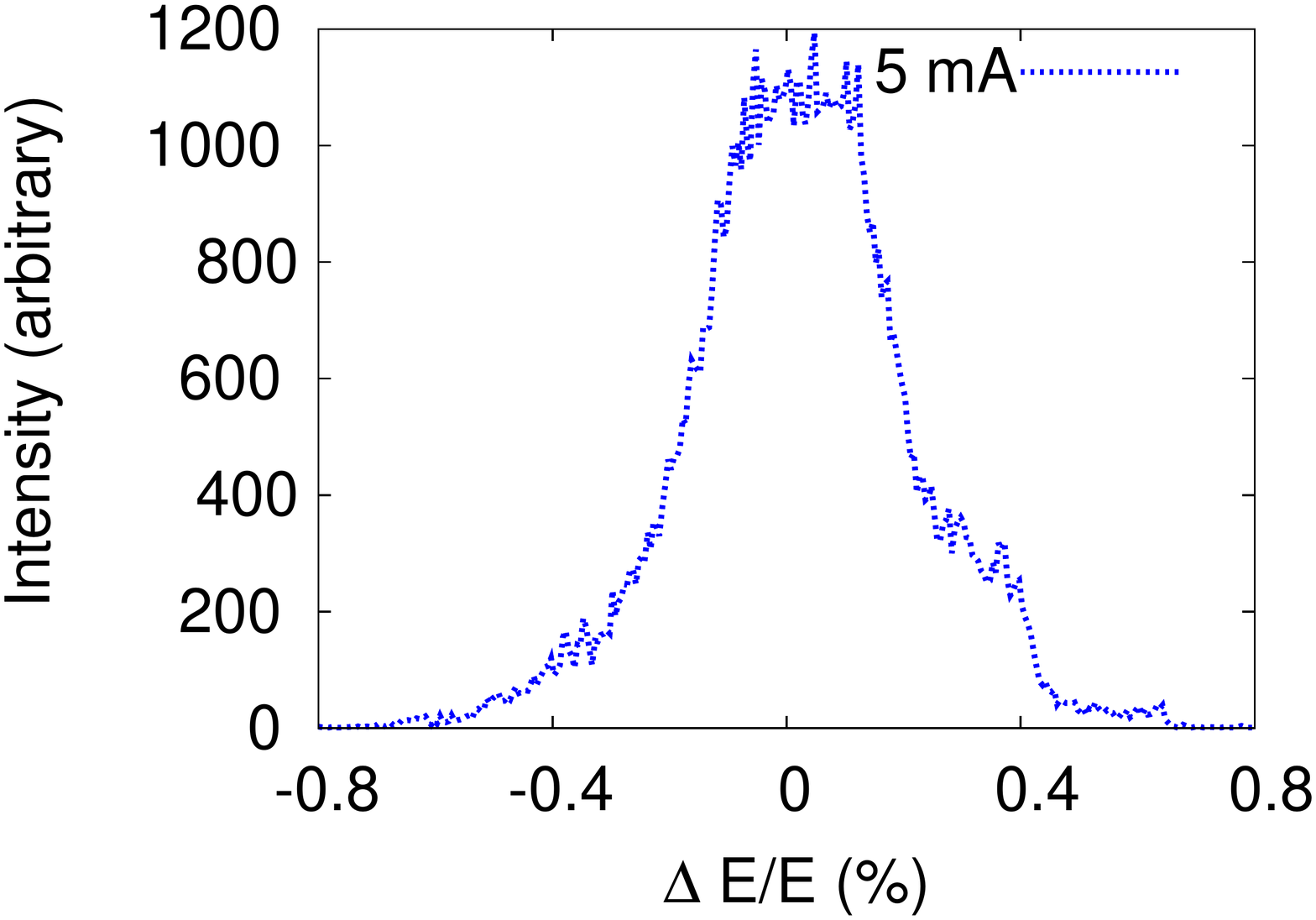}}
\end{center}
\caption{ The radial  profile and energy-spread-histogram of the \htp\ 5 mA beam on the stripper.   
\label{fig:ch2-2} }
\end{figure}

For cyclotrons using a multi-turn extraction scheme, obtaining the particle distribution at the stripper is critical because it
constitutes the initial conditions for the subsequent beam transport in the extraction channel.
For the DSRC cyclotron  this is especially important because the protons in the extraction channel have the same sign of charge but a double charge-to-mass ratio than the \htp\ particle. Protons will travel inwards guided by the complicated magnet
and rf-fields, as is shown in Fig.\ \ref{fig:ch2-3}. In Fig.\ \ref{fig:ch2-2} we shows the radial profile and energy spread on the stripper. 
In this initial study, we neglect scattering on the foil itself, as well as the dynamics of the stripped electrons. 
For the 5 mA beam we find that the energy spread stays at $\pm$0.6\% at a phase width of 12$^\circ$. 
The transverse normalized rms emmitance are  $\epsilon_{r,n}$=1.1 mm-mrad and $\epsilon_{z,n}$=1.4 mm-mrad.
We take this result as the initial distribution for proton extraction simulation, which  shows no problem for the beam transport in the extraction channel. 

Detailed discussion of the space-charge effects in the DSRC cyclotron will be addressed in a forthcoming paper\ \cite{yang-2}.


\section{Radio frequency design}
Focusing at the problem of vacuum and rf-voltage leakage at very high electromagnetic fields, we pose the question if the use of $8$ double gap rf-cavities could be advantageous over a  solution with $4$ single gap rf-cavities.
The double gap rf-cavities have the following advantages:
\begin{itemize}
  \item It is possible to install a cryopanel inside the electrodes of the double gap cavities, near to the median plane and well shielded from the beam (4-10 cm far from the median plane). In consequence we have a large cryopanel in each valley.
\item The maximum voltage could stay around 300 kV; hence, the problem of the rf-stray fields should be significantly reduced.
\item The accelerating voltage is more uniform vs. orbit radius and of 4000 keV/turn at all the radii, while if we have only 4 single gap cavities we have a maximum of 4 MV just in the range 4.0 - 5.0 m.
\end{itemize}


\subsection{DIC rf-design}
\begin{figure}[H]
  \centering
  \includegraphics[width=0.5\textwidth]{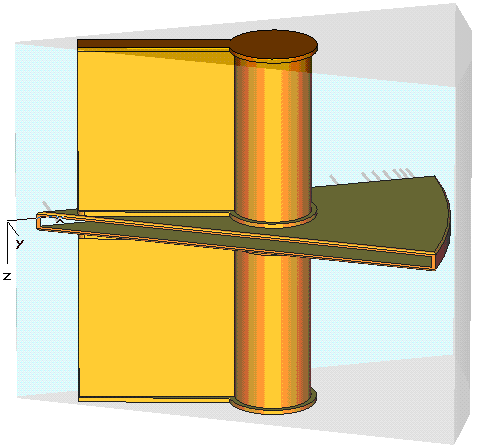}
  \caption{CAD geometry for the DIC rf-cavity}
  \label{fig:Injector_cavity}
\end{figure}

\begin{figure}[H]
  \centering
  \includegraphics[width=0.85\textwidth]{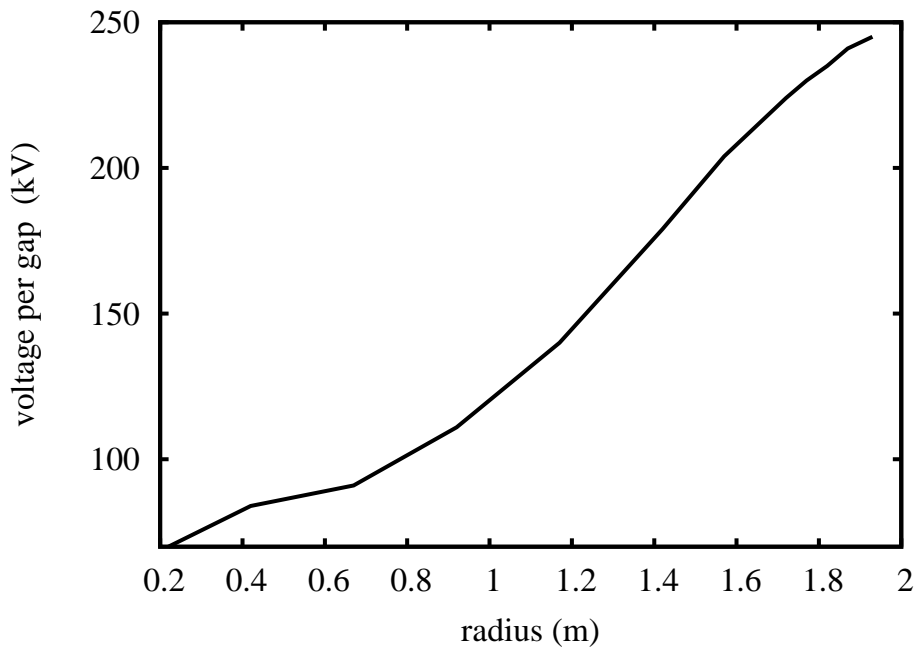}
  \caption{Gap voltage distribution in the DIC rf-cavity as function of the cyclotron radius.}
  \label{fig:INJ_cavity_Vgap}
\end{figure}

The rf-cavity system of the DAE$\delta$ALUS Injector Cyclotron consists of four double-gap cavities  ($\lambda/2$ resonators) with one stem, that produce an energy gain of ~0.5 MeV per cavity at the extraction radii (1900 mm) allowing the acceleration of the beam up to 60 MeV/amu.
Double-gap cavities are suitable for applications in which a special radial voltage profile (along the acceleration gaps as shown in Fig.\ \ref{fig:INJ_cavity_Vgap}) is desired. The rf-voltage must not exceed 400 kV per gap, limiting the acceleration performance and requiring a complex cooling system. The double-gap cavity solution was modeled using the multi stem approach~\cite{RF-REF1}, the CAD-geometry is shown in Fig.\ \ref{fig:Injector_cavity}. Double-gap cavities can be easily installed in the valley, they allow the insertion of cryo-pumps or cryo-panels adjacent to the median plane, an advantage with respect to increasing the performance of the vacuum system. Electrostatic deflectors can be installed at the extraction radius inside the dee if necessary, which shape will provide the required space. 

\begin{table}[H]
  \centering
   \caption{Parameters and performances of the DIC double-gap cavity }
  \begin{tabular} {lrrl}
  \hline
   & Injector Cavity & Unit\\
  \hline
  Cavity Height	& 1700 &  mm \\
Cavity Angular Extension (Width)	&  32 & deg \\
Cavity Radial Extension (Length)	&  2100 & mm \\
Dee Angular Extension (Width)	&  30 & deg \\
Dee Gap	& 50 & mm \\
Dee Thickness	& 20 & mm \\
Stems Radial Extension	& 320-1330 & mm \\
Stems Angular Extension	& 70 & mm \\
Resonance mode   	& double-gap, $\lambda/2$  & \\
Resonance frequency	& 49.2 & MHz \\
Quality factor	 & 9100 & \\
rf-Power Dissipation	 & 160 & kW \\
Energy Gain (voltage distribution see Fig.\ \ref{fig:INJ_cavity_Vgap}) 	& 70-250 & kV \\
Max Surface Current	& 160 & A/cm \\
Max Electric Field 	& 6.1 & MV/m\\
    \hline
  \end{tabular}
   \label{RF_Injector_Cavity}
\end{table}
The cavity is a resonator with one stem that is connected to the liner and the dee by means of a system of flanges. The shape of the dee and the shape of the stem were optimized to fit the space requirements and to satisfy the rf-specifications. 
Table~\ref{RF_Injector_Cavity} lists the main parameters of the injector double-gap cavity.

Rf-power will be fed to the cavity via a capacitive coupler, frequency tuning pistons will be located in the same area.

\subsection{DSRC rf-design}

The rf-cavity system of the DAE$\delta$ALUS Superconducting Ring Cyclotron consists of four single-gap cavities that are similar to the PSI box shaped cavities and produce an energy gain of ~4 MeV per turn at extraction radii (4900 mm). Because of the large energy gain and the resulting decrease in acceleration time, the beam losses due to the interaction with the residual gases are minimized. Double-gap cavities (multistem $\lambda/2$ resonators) represent a valid option and are sketched briefly.

\subsubsection{Double-gap cavities}
Unfortunately it is not practical to install more than four single-gap cavities in an 8-sector cyclotron due to the mechanical interference of the cavities in the central region and the space needed for other elements. Therefore, despite the double gap cavities being less efficient than the single gap cavities, two double gap cavities can be added to increase the energy gain per turn by an additional 0.8 MeV, thus increasing the maximum beam power of about 20$\%$, without increasing beam losses.
\subsubsection{Single-gap cavities}
\begin{figure}[H]
  \centering
  \includegraphics[width=0.454\textwidth]{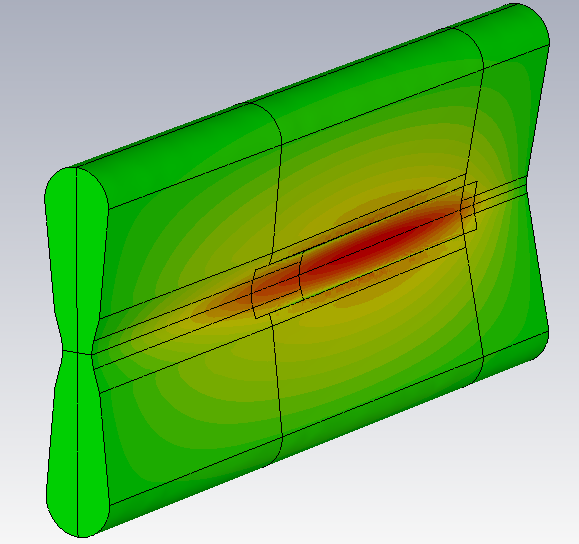}\includegraphics[width=0.5\textwidth]{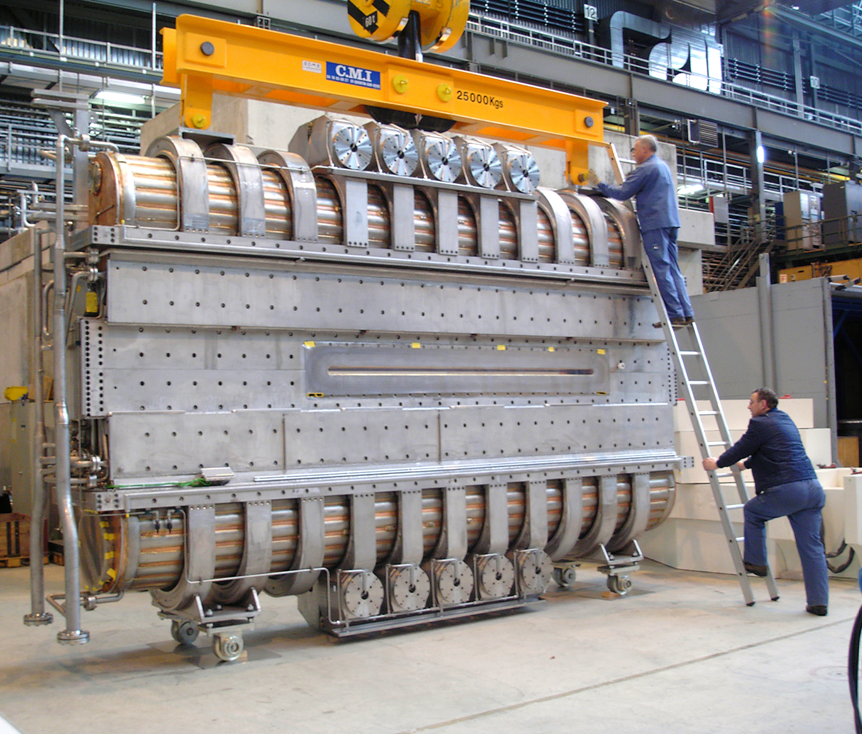}
  \caption{Electric-field distribution in the DSRC cavity (left) and existing PSI cavity (right).}
  \label{DSRC_cavity}
\end{figure}

Single-gap cavities can reach higher accelerating voltages (with sinusoidal distribution), and are less sensitive to higher-order modes. The cooling system, braised onto the outside of the cavity, is simple to fabricate with no risk of water leaking into the vacuum (high reliability)~\cite{RF-REF2}. Engineering complexity and bigger dimensions are issues, but proven to be manageable, as shown at PSI.
A single-gap cavity with a gradually modulated section along the radius was designed and optimized to fulfil the performance requirements. The cavity wall consists of an 8 mm copper sheet on which cooling channels are directly tungsten inert gas (TIG) brazed. A large number of channels provides  efficient water-cooling with small thermal gradients. The total height of the cavity is 3000 mm. The cavity radial extensions is from radius 900 mm to radius 7700 mm while the acceleration gap ranges from 150 mm to 300 mm (the acceleration is realized from radius 1900 mm to 4900 mm). The power will be fed into the cavity by two high power inductive couplers. 
A stainless steel support structure is required to provide mechanical stability to the cavity. The tuning of the cavity will be performed by means of the trimming system currently used at PSI \cite{RF-REF3}. This complex but effective cavity tuning system is realized by hydraulic tuning yoke that squeezes the cavity to compensate the frequency variation.
No parallel surfaces represent a valid solution to avoid multipactoring.
Four cryopumps located two per side of the cavity will provide the required $10^{-8}$mbar vacuum. The vacuum between the cavity and the vacuum chamber of the coils is guarantee by a system based on inflating seals and by o-rings; this technology has been successfully developed and optimized at PSI and RIKEN \cite{psi-cycl-2010}.
Figure~\ref{DSRC_cavity} shows a 3D model of the cavity,  in Table~\ref{RF_DSRC_Cavity} the main parameters of the DSRC single-gap cavity are compared to the existing PSI single-gap cavity.

\begin{table}[htbp]
  \centering
   \caption{Comparison of the rf-requirements for DSRC and achieved performance at the PSI 590 MeV Ring Cyclotron.}
  \begin{tabular}{lrrl}
  \hline
   & \textbf{DSRC} & \textbf{PSI 590 MeV RC} & Unit\\
  \hline
Cavity height    & 3000  & 2928 & mm\\
Cavity length    & 6800 & 5200 & mm\\
Cavity opening in the beam plane (injection)    & 150 & 300 & mm\\
Cavity opening in the beam plane (extraction)    & 300 & 300 & mm\\
Cavity copper sheet thickness    & 8 & 8 & mm\\
Material of rf-surface    & OFHC Copper & OFHC Copper & \\
Material of support structure & 316LN & 316LN & \\
Resonance mode    & TM011& TM011 & \\
Resonance frequency     & 49.2 & 50.6 & MHz\\
Accelerating voltage per turn at the inner radii     & 2.0 & 3.4 & MV\\
Accelerating voltage per turn at the outer radii     & 4.0 & 4.2 & MV\\
Voltage distribution on a gap    & 0.5-1.0 & half-sine & MV \\
Amplitude stability & & 0.03 & $\%$ \\
Phase stability & & 0.01 & deg\\
Quality factor      & 37,500 & 44,200\\
rf-Power Dissipation for 1 MV gap voltage & 500 & 300 & kW\\
Wall plug power AC/DC conversion efficiency & 0.9 & 0.9 & \\
DC / rf conversion  efficiency & 0.64 & 0.64 & \\
rf / average beam-power conversion efficiency & 0.67& 0.55 & \\
  \hline
  \end{tabular}
  \label{RF_DSRC_Cavity}
\end{table}

\begin{figure}[H]
  \centering
  \includegraphics[width=0.85\textwidth]{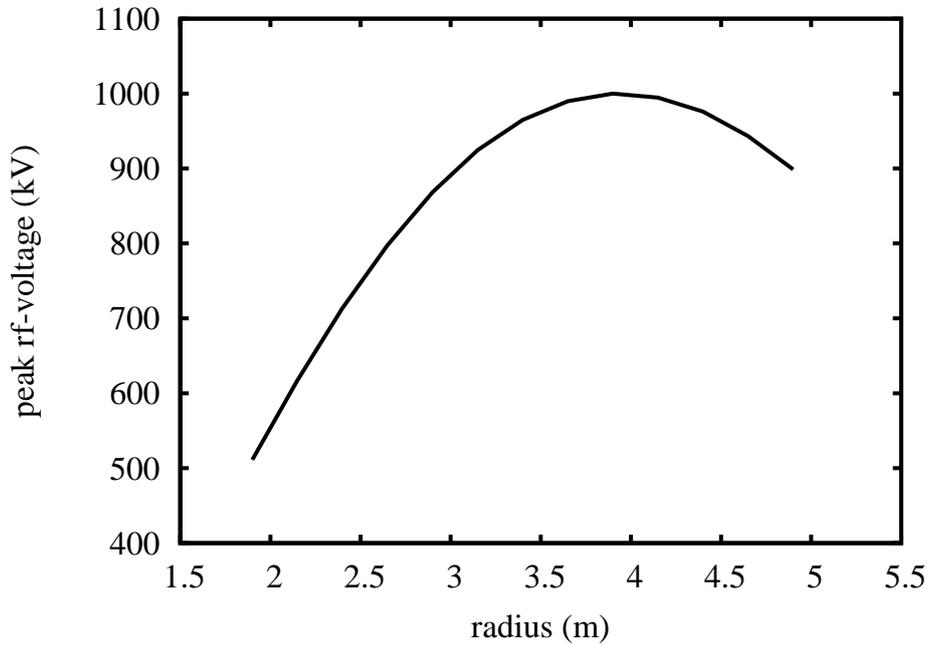}
  \caption{Gap voltage distribution in the DSRC cavity as function of the cyclotron radius.}
  \label{DSRC_cavity_Vgap}
\end{figure}

The DSRC resonator works at the sixth harmonic (resonant frequency of 49.2 MHz) with a quality factor (Q value) of 37,500, the gap voltage distribution is shown in Fig.\ \ref{DSRC_cavity_Vgap}. The cavity needs 0.5 MW of power to produce an accelerating voltage that ranges from 0.5 MV at the inner radii to 1 MV at the extraction radius. The acceleration gap at the inner radii is reduced from 300 mm to 150 mm to fit into the restricted space of the valley due to the cryostat of the superconducting coils.

\subsubsection{Duty cycle}
\begin{figure}[H]
  \centering
  \includegraphics[width=0.85\textwidth]{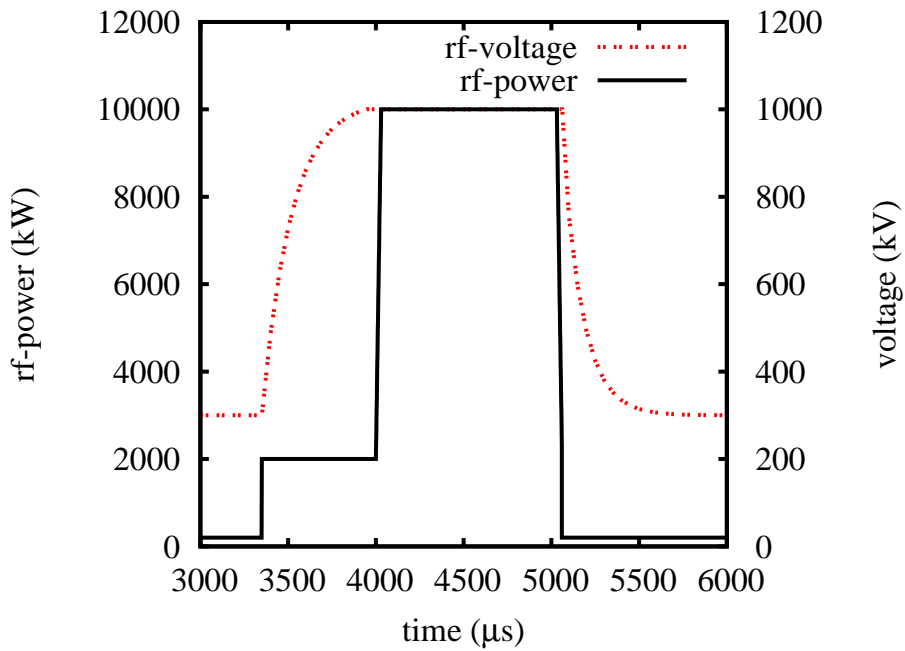}
  \caption{Example of a typical time-structure applicable for the DSRC rf-pulse.}
  \label{RF-PulseStructure}
\end{figure}

One of the challenges of the rf-system is the pulsed structure of the beam that is required for the final application, as shown in Fig.\ \ref{RF-PulseStructure}.
The beam must be turned ON 20$\%$ of the time but the pulsing frequency could be selected in order to minimize the stress on the amplifier chains as well as the cavity voltage regulation. Another parameter to take into account is to minimize the global power consumption of the installation. For that it could be interesting to reduce the cavity voltage during the beam OFF times. Different scenarii have been analyzed from pulse lengths of 100 $\mu s$ to a few seconds. The high Q value of the cavity defines a time constant of more than 200 $\mu s$. In consequence, the filling time of the cavity would be at least 500 $\mu s$ considering some overdriving. This probably rejects a pulse length of 100 $\mu s$ or shorter.
A pulse duration on the order of 1ms would allow us to reduce the rf-voltage during 3/5th of the time. In this case the mean rf-power is reduced by 1.2 MW. This translates in a plug power reduction of nearly 2 MW, considering the efficiency of the rf-amplifiers. A pulse of 10 ms would further reduce the power consumption by 600 kW. An additional advantage is that the cavity cooling is greatly relaxed.
Between 1 ms and 10 ms pulse duration, the tube power supplies can rely on their output filter bank to provide the peak power. The power supply can then be dimensioned to deliver only the average power.

One important point could also be related to the capability of the cavity voltage regulation to compensate fast enough for the beam loading.

The time for the beam to make the full acceleration in the ring cyclotron is about 30\ $\mu s$. This means that when the beam will be turned ON-OFF the required rf-power will rise-fall linearly during this time. This will naturally help the regulation loop to follow.
In order to help the negative feed-back loop during the transients a feed-forward signal could be injected in order to anticipate the driving power requirement. Some examples of electrons accelerators  with very large beam loadings have demonstrated very good voltage stability of a few $10^{-4}$ (for example IBA Rhodotron with beam loading of up to 7:1) during the transients using this technique. 
Pulse lengths of more than 10\ ms can start creating problems of very high energy stored in the power supply in order to be able to deliver the peak power. A 10\ ms pulse means a pulsing frequency of 20\ Hz. At these levels we enter into the mechanical resonances region of the delicate grid structures of the amplifier tubes.
Pulse length of 100\ ms and more can create additional issues with electricity grid loading. Main parts of the power supplies should then be dimensioned for the maximal power. This would have an important cost impact.
From the pulse structure of the $H_2^+$ beam and the expected performance of the accelerating cavities, a total peak-power of 2.5 MW during beam extraction, and a 600 kW average power has to be fed to each cavity.

The heavy beam loading induces a large swing of the cavity input impedance. If the cavity is matched ($50~\Omega$ input impedance) during beam acceleration, the amplifier and transmission line have to deal with an input impedance of about $250~\Omega$ when the beam is off. This leads to a Voltage Standing Wave Ratio of $VSWR \approx 5$ and a relative voltage enhancement of about 67\% in the transmission line.

\begin{figure}[htbp]
  \centering
  \includegraphics[width=1.0\textwidth]{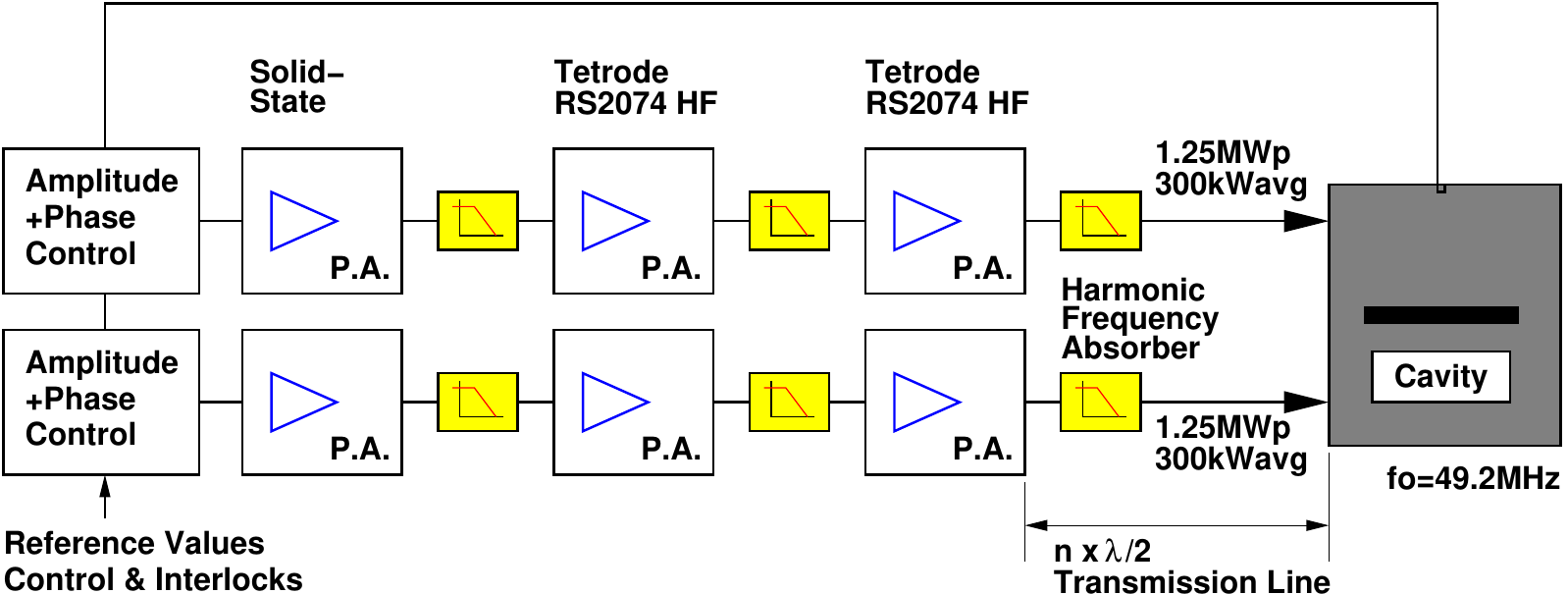}
  \caption{Proposed MW-class amplifier chain}
  \label{fig:pa-layout}
\end{figure}

\begin{table}[htbp]
  \centering
    \caption{Comparison of the rf-requirement for the DSRC and achieved performance at PSI}
  \begin{tabular} {lrrl}
  \hline
   & DSRC & PSI 590 MeV RC & Unit\\
  \hline
  Operating Frequency & 49.2 & 50.6 & MHz \\
  Cavity gap voltage & 1 & 1.4 & MV \\
  Average cavity input power & 600 & 700 & kW \\
  Maximum amplifier power & 2.5 peak & 1 cw & MW \\
  Beam loading factor & 5.0 & 1.2 & \\
  \hline
  \end{tabular}
  \label{RF_Parameters}
\end{table}
Since the requirements of the DSRC are close to the achieved performance of the PSI rf-system a possible feeding scheme of the cavities can be extrapolated from the PSI high power amplifier chain. In order to reduce the peak power of the final amplifier stage, two MW-class amplifier chains could be used and coupled to the cavity, as illustrated in Fig.\ \ref{fig:pa-layout}. Each cavity will then be equipped with two input power coupler and amplifier chains.

A more modern and powerful power tube for the final amplifier would be the TH525 tetrode, for example. With this tetrode, it might be possible to use one amplifier chain per cavity only. Alternative  final amplifier tubes are the  4CM2500KG and 8973 from CPI International.

\section{Comments on the scope of this article}

The scope of this article has been limited to the systems in the
accelerator chain that are critical for the high-intensity design of
DAE$\delta$ALUS.  These topics are expected to be of wide interest to
readers.  Several, mostly specialized, topics (subsystems) have not been considered, and we
touch on these here briefly.

We have omitted subsystems which 
 are not critical from the point of view of high-intensity operation 
and which are standard for
 many operating accelerator facilities. 
 Such systems include beam diagnostics and controls, the
 beam transfer lines between injector and ring cyclotron, and the
 transfer line to the beam target.    

This article has also not included in-depth discussion of the beam 
targeting.   Development of the target station is an iterative
process that accounts for the physics demands for the neutrino flux as
well as the operational demands of the high-power beam on the target.   The
design is underway, inspired by target stations at LANL, ISIS, and PSI.
Details of the target station and the neutrino flux will be given
in a future article.  

However, we point out a few facts about the targeting here.   
The beam target must accept high average power, but has no special
 constraints in view of the power density.  It is possible to use
 copper as a material with high thermal conductivity.  The beam size
 can be beam-optically enlarged to obtain a power density that can be
 safely handled in the copper. PSI operates a MW-class beam dump, made
 from copper with a conical shape to distribute the beam power in
 the longitudinal direction. The dump is sectioned into three 0.5\,m long segments, to allow a simple exchange of the highly activated segments
 in case of failure.  This dump can serve as a model for higher beam
 powers. The dimensions can be scaled up to keep the peak temperatures
 at acceptable levels.

 Conventional subsystems are needed. These include the
 buildings, concrete and steel shielding, electrical installations,
 radiation monitoring, interlock systems, and venting systems, including
 monitoring of mobile radioactive isotopes. Practically, the complete
 electrical power which is used to operate the facility and to
 generate the intense beam has to be removed from the components by
 water-cooling circuits. Significant installations will be required
 for the conventional cooling systems in this high-power accelerator.
 Finally, specific auxiliary systems are needed for the operation of
 high-intensity accelerators and targets. Such systems are, for example,
 shielded exchange flasks for activated components and a hot cell
 facility for repair work. Much of the design of the conventional subsystems will be site-specific
 and designed in a later stage of DAE$\delta$ALUS development.

\section{Conclusions}

In this paper we have addressed some of the most challenging research questions facing the DAE$\delta$ALUS project regarding a cyclotron-based high-power proton driver in the megawatt range with a kinetic energy of 800 MeV.  Cyclotrons of this kind would be useful for numerous applications, ranging from particle physics to nuclear waste transmutation.

We identified \htp\  as the preferred particle to accelerate, as opposed to bare protons, on the basis of beam dynamics arguments which show that substantially higher beam currents are possible.  The area found to be most challenging, and requiring greater research efforts, was the discussion about suppressing \htp\  ions with higher-order vibrational states and maintaining the general loss budget at required order of $10^{-4}$ of the total intensity.

Precise beam dynamics simulations with 3D space charge and extraction using a simple \htp\  stripping process are the bases for characterization and quantification of the beam halo -- one of the most limiting processes in high-power particle accelerators.  We show that in the DIC controlled beam losses are well within the allowed margins, and the stripped proton beam from the DSRC can be well matched to the extraction channel.

Initial calculations of vacuum and dissociation cross sections are also within the expectations;  however, more dedicated research is needed.  For this purpose, a two-phased program is being initiated to optimize source performance and inflection/capture into the DIC.  The first phase is an experiment for inflection and capture into the central region of the DIC; the second is a study of vibrational state distributions in the beam and methods of suppressing the weakly bound states.

The initial design of the rf-system, mostly based on enhanced versions of the PSI cavities, is shown to be feasible for delivering the required rf-power.  The required pulse structure (20\% duty cycle) is expected to be feasible based on the presented initial calculations.

Achieving the very high proton currents and the required power on target is based on generalized perveance arguments:  the space-charge forces at injection energies for the required currents of \htp\  closely match the conditions in existing high-current proton machines with substantially less beam current.  Although the design of the DSRC and the DIC certainly requires further refinement, the design example presented demonstrates the feasibility of achieving the very high performance goals needed for the DAE$\delta$ALUS accelerators.

\section*{Acknowledgments}
Support for this workshop has been provided through the Majorana Centre from the INFN Eloisatron Project, directed by Prof. Antonino Zichichi.
Support for studies is provided by the National Science Foundation and the Massachusetts Institute of Technology.
The majority of computations have been performed on the Cray XT4 in the
framework of the PSI CSCS ``Horizon'' collaboration and the local PSI computing resources FELSIM and Merlin4.  
\bibliography{Erice-2011-bib}
\bibliographystyle{elsarticle-num}

\end{document}